\documentclass[symmetry,article,accept,moreauthors,dvipdfm,12pt,a4paper]{mdpi} 
 \lastpage{xxx} \doinum{10.3390/xxxx}
\pubvolume{xx} \pubyear{2009} \history{Received: 23 October 2009 /
Accepted: 13 November 2009 / Published: xx}

\usepackage{amssymb,amsmath}
\usepackage{graphicx}
\usepackage{epsfig}
\usepackage{color}
\usepackage{caption}
\usepackage{hyphenat}
\usepackage{hyperref}
\hypersetup{
    colorlinks,%
    citecolor=black,%
    filecolor=black,%
    linkcolor=black,%
    urlcolor=black
}

\def\paap{\,_c\left\langle 11 \mid 1'1' \right\rangle_c}
\def\ka{\left.\mid 11 \right\rangle_c}
\def\kap{\left.\mid 1'1' \right\rangle_c}
\def\kb{\left.\mid 88 \right\rangle_c}
\newcommand{\bq}{\bar q}
\newcommand{\bn}{\bar n}
\newcommand{\bs}{\bar s}

\newcommand{\br}{\left<}
\newcommand{\et}{\right>}
\newcommand{\eref}[1]{(\ref*{#1})}
\newcommand{\ktt}{|\bar 33\rangle_c^{12}}
\newcommand{\kss}{|6\bar 6\rangle_c^{12}}
\newcommand{\btt}{^{12}_c\langle\bar 33|}
\newcommand{\bss}{^{12}_c\langle 6\bar 6|}
\def\kbp{\left.\mid 8'8' \right\rangle_c}
\def\ba{\,_c\left\langle 11 \mid\right.}
\def\bb{\,_c\left\langle 88 \mid\right.}
\def\bap{\,_c\left\langle 1'1' \mid\right.}
\def\bbp{\,_c\left\langle 8'8' \mid\right.}

\title{Tetraquark Spectroscopy: A Symmetry Analysis}

\author{Javier Vijande $^{1,*}$ and Alfredo Valcarce $^{2}$}

\address{$^{1}$ Departamento de F\'{\i}sica At\'{o}mica, Molecular y Nuclear, Universidad de
Valencia (UV) and IFIC (UV-CSIC), Valencia, Spain\\
$^{2}$ Departamento de F\'{\i}sica Fundamental, Universidad de
Salamanca, Salamanca, Spain; E-Mail:~valcarce@usal.es}

\emails{javier.vijande@uv.es; valcarce@usal.es}

\corres{javier.vijande@uv.es, Tel: +34 963543883}
\abstract{We present a detailed analysis of the symmetry properties
of a four-quark wave function and its solution by means of a
variational approach for simple Hamiltonians. We discuss several
examples in the light and heavy-light meson sector.}


\begin{document}


\section{Introduction}
The potentiality of the quark model for hadron physics in the
low-energy regime became first manifest when it was used to classify
the known hadron states. Describing hadrons as $q\bq$ or $qqq$
configurations, their quantum numbers were correctly explained. This
assignment was based on the comment by Gell-Mann~\cite*{Ge64}
introducing the notion of quark: {\it ``It is assuming that the
lowest baryon configuration ($qqq$) gives just the representations
1, 8 and 10, that have been observed, while the lowest meson
configuration ($q \bar{q}$) similarly gives just 1~and~8''}. Since
then, it has been assumed that these are the only two configurations
involved in the description of physical hadrons. However, color
confinement is also compatible with other multiquark structures like
the tetraquark $qq\bq\bq$ first introduced by Jaffe \cite{Ja77}.
During the last two decades there appeared a number of experimental
data that are hardly accommodated in the traditional scheme defined
by Gell-Mann.

One of the first scenarios where the existence of bound multiquarks
was proposed was a system composed of two light quarks and two heavy
antiquarks ($nn\bar Q\bar Q$). These objects are called heavy-light
tetraquarks due to the similarity of their structure with the
heavy-light mesons ($n\bar Q$). Although they may be experimentally
difficult to produce and also to detect~\cite{Mo96} it has been
argued that for sufficiently large heavy quark mass the tetraquark
should be bound~\cite{ZS86a,ZS86b}. The stability of a heavy-light
tetraquark relies on the heavy quark mass. The heavier the quark the
more effective the short-range Coulomb attraction to generate
binding, in such a way that it could play a decisive role to bind
the system. Moreover the $\bar Q \bar Q$ pair brings a small kinetic
energy into the system contributing to stabilize it.

\begin{figure}[h]
  \centering
  \caption{Quark content of a $q\bq$ nonet (left) and a $qq\bq\bq$ nonet (right).}
  \includegraphics{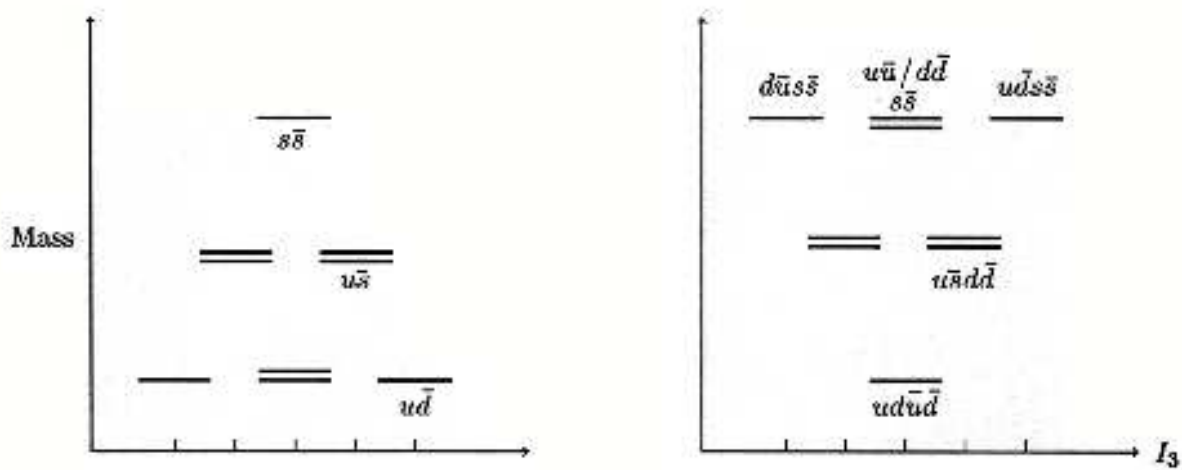}
  \label{fig-jaff}
  \end{figure}

Another interesting scenario where tetraquarks may be present corresponds to the scalar mesons, $J^{PC}=0^{++}$.
To obtain a positive parity state from a $q\bq$
pair one needs at least one unit of orbital angular momentum. Apparently
this costs an energy around 0.5 GeV\footnote{This
effect can be estimated from the experimental $M(L=1)-M(L=0)$ energy differences:
$a_1(1260)-\rho(776)=484$ MeV, $f_1(1282)-\omega(782)=500$ MeV,
$h_1(1170)-\eta(548)=622$ MeV, $h_c(3526)-\eta_c(2980)=546$ MeV,
$\chi_{c1}(3511)-J/\Psi(3097)=414$ MeV, $\chi_{b1}(9893)-\Upsilon(9460)=433$ MeV,
being the average $M(L=1)-M(L=0)\approx500$ MeV.}, making the lightest theoretical scalar
states to be around 1.3 GeV, far from their experimental error bars. However, a $qq\bq\bq$ state
can couple to $J^{P C}=0^{++}$ without orbital excitation and, as a consequence, they could coexist and mix
with $q\bq$ states in this energy region.  Furthermore, the color and spin
dependent interaction arising from the one-gluon exchange, favors states where quarks and
antiquarks are separately antisymmetric in flavor. Thus, the energetically
favored flavor configuration for $qq\bq\bq$ is $[(qq)_{\bar 3}(\bq\bq)_3]$, a
flavor nonet, having the lightest multiplet spin 0. The most striking feature
of a scalar $qq\bq\bq$ nonet in comparison with a $q\bq$ nonet is a {\it reversed
mass spectrum} (see Figure~\ref*{fig-jaff}). One can see a degenerate isosinglet
and isotriplet at the top of the multiplet, an isosinglet at the bottom, and a
strange isodoublet in between. The resemblance to the experimental structure of the
light scalar mesons is striking.

Four-quark states could also play an important role in the charm
sector. Since 2003 there have been discovered several open-charm
mesons: the $D_{sJ}^*(2317)$, the $D_{sJ}(2460)$, and the
$D_0^*(2308)$. In the subsequent years several new states joined
this exclusive group either in the open-charm sector: the
$D_{sJ}(2860)$, or in the charmonium spectra: the $X(3872)$, the
$X(3940)$, the $Y(3940)$, the $Z(3940)$, the $Y(4260)$, and the
$Z(4430)$ among others~\cite{PDG08}. It seems nowadays unavoidable
to resort to higher order Fock space components to tame the
bewildering landscape arising with these new findings. Four-quark
components, either pure or mixed with $q\bar q$ states, constitute a
natural explanation for the proliferation of new meson
states~\cite{Jaf05a,Jaf05b,Jaf05c}. They would also account for the possible
existence of exotic mesons as could be stable $cc\bar n\bar n$
states, the topic for discussion since the early 1980s~\cite{Ade82a,Ade82b}.

All these scenarios suggest the study of $qq\bq\bq$ structures and
their possible mixing with the $q\bq$ systems to understand the role
played by multiquarks in the hadron spectra. The manuscript is
organized as follows. In Section~\ref*{tech} the variational
formalism necessary to evaluate four-quark states is discussed in
detail with special emphasis on the symmetry properties. In Section
\ref*{thres} the way to exploit discrete symmetries to determine the
four-quark decay threshold is discussed. In Section \ref*{prob} the
formalism to evaluate four-quark state probabilities is sketched.
In Section~\ref*{results} we discuss some examples of four-quark
states calculated using this formalism. Finally, we summarize in
Section~\ref*{summary} our conclusions.

\section{Four-quark spectra}
\label{tech}

\subsection{Solving the four-body system}
\label{mixo}

The four-quark ($qq\bq\bq$) problem will be addressed by means of
the variational method, specially suited for studying low-lying
states. The nonrelativistic Hamiltonian will be given by
\begin{equation}
H=\sum_{i=1}^4\left(m_{i}+\frac{\vec p_{i}^{\,2}}{2m_{i}}\right)+\sum_{i<j=1}^4V(\vec r_{ij}) \, ,
\label{ham}
\end{equation}
where the potential $V(\vec r_{ij})$ corresponds to an arbitrary
two-body interaction. The extension of this formalism to consider
many-body interactions is discussed in~\cite{Vij07ba,Vij07bb}.

The variational wave function must include all possible flavor-spin-color channels
contributing to a given configuration. For each channel $s$, the wave function will be the tensor product of
a color ($\left|C_{s_1}\right>$), spin ($\left|S_{s_2}\right>$), flavor ($\left|F_{s_3}\right>$), and radial
($\left|R_{s_4}\right>$) component,
\begin{equation}
\label{efr}
\left| \phi _{s}\right>=\left|C_{s_1}\right>\otimes\left|
S_{s_2}\right>\otimes\left|F_{s_3}\right>\otimes\left|R_{s_4}\right> \, ,
\end{equation}
where $s\equiv\{s_1,s_2,s_3,s_4\}$.  The procedure to construct the wave function will be detailed later
on. Once the spin, color and flavor parts are integrated out the coefficients of the radial wave function are
obtained by solving the system of linear equations
\begin{equation}
\label{funci1g}
\sum_{s'\,s} \sum_{i} \beta_{s_4}^{(i)}
\, [\langle R_{s_4'}^{(j)}|\,H\,|R_{s_4}^{(i)}
\rangle - E\,\langle
R_{s_4'}^{(j)}|R_{s_4}^{(i)}\rangle \delta_{s,s'} ] = 0
\qquad \qquad \forall \, j\, ,
\end{equation}
where the eigenvalues are obtained by a minimization procedure.

\subsection{Four-body wave function}
\label{tetwav}

\begin{figure}[tb]
\begin{center}
\caption[Tetraquark Jacobi coordinates.]{Tetraquark Jacobi
coordinates, see Equations~\eref{coo} for definitions.} \label{coor}
\epsfig{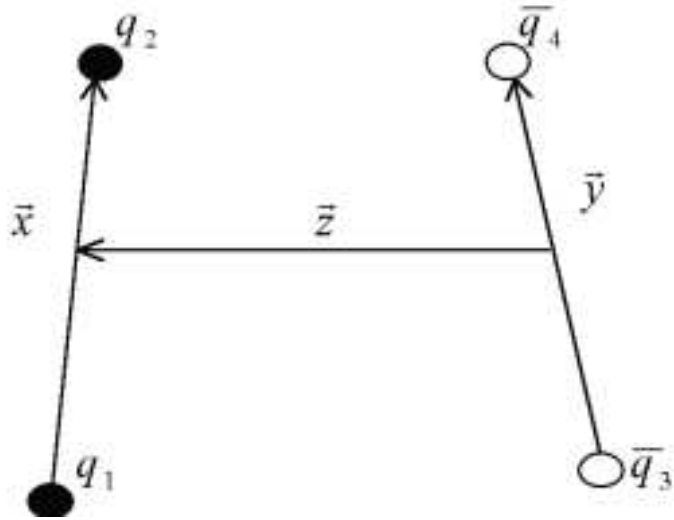}
\end{center}
\end{figure}

For the description of the $q_1q_2\bar{q_3}\bar{q_4}$ wave function
we consider the four-body Jacobi coordinates depicted in
Figure~\ref*{coor}:
\begin{eqnarray}
\label{coo}
\vec{x} &=&\vec{r}_{1}-\vec{r}_{2} \\ \nonumber
\vec{y} &=&\vec{r}_{3}-\vec{r}_{4} \\ \nonumber
\vec{z} &=&\frac{m_{1}\vec{r}_{1}+m_{2}\vec{r}_{2}}{m_{1}+m_{2}}-\frac{m_{3}\vec{r}_{3}+m_{4}\vec{r
}_{4}}{m_{3}+m_{4}}\\ \nonumber
\vec{R} &=&\frac{\sum m_{i}\vec{r}_{i}}{\sum m_{i}}\nonumber \, ,
\end{eqnarray}

\noindent where indices $1$ and $2$ will stand for quarks and $3$ and $4$ for antiquarks.
Let us now describe each component of the variational wave function \eref{efr} separately.
The total wave function should have well-defined permutation properties under
the exchange of identical particles: quarks or antiquarks. The Pauli principle
must be satisfied for each subsystem of identical
particles\footnote{One should have in mind that if flavor $SU(3)$ symmetry
is assumed, $u$, $d$, and $s$ quarks are identical particles.}.
This imposes restrictions on the quantum numbers of the basis states.

\subsection{Color space}

There are three different ways of coupling two quarks and two antiquarks into a colorless state:
\begin{subequations}
\begin{eqnarray}
\label{eq1a}
[(q_1q_2)(\bar q_3\bar q_4)]&\equiv&\{|\bar 3_{12}3_{34}\rangle,|6_{12}\bar 6_{34}\rangle\}\equiv\{|\bar 33\rangle_c^{12},
|6\bar 6\rangle_c^{12}\}\\
\label{eq1b}
[(q_1\bar q_3)(q_2\bar q_4)]&\equiv&\{|1_{13}1_{24}\rangle,|8_{13} 8_{24}\rangle\}\equiv\{|11\rangle_c,|88\rangle_c\}\\
\label{eq1c}
[(q_1\bar q_4)(q_2\bar q_3)]&\equiv&\{|1_{14}1_{23}\rangle,|8_{14} 8_{23}\rangle\}\equiv\{|1'1'\rangle_c,|8'8'\rangle_c\}\,,
\end{eqnarray}
\label{eq1}
\end{subequations}

\noindent being the three of them orthonormal basis. Each coupling
scheme allows to define a color basis where the four-quark problem
can be solved. Only two of these states have well defined
permutation properties: $\ktt$, is antisymmetric under the exchange
of both quarks and antiquarks, $(AA)$, and $\kss$ is symmetric,
$(SS)$. Therefore, the basis Equation~(\ref*{eq1a}) is the most
suitable one to deal with the Pauli principle. The other two,
Equations~(\ref*{eq1b}) and~(\ref*{eq1c}), are hybrid bases
containing singlet-singlet (physical) and octet-octet (hidden-color)
vectors. The three basis are related through~\cite{De63a,De63b}:
\begin{eqnarray}
\label{qq13}
|11\rangle_c&=&\sqrt{1\over3}\,\ktt+\sqrt{2\over3}\,\kss\\ \nonumber
|88\rangle_c&=&-\sqrt{2\over3}\,\ktt+\sqrt{1\over3}\,\kss \, ,
\end{eqnarray}
and
\begin{eqnarray}
\label{qq14}
|1'1'\rangle_c&=&-\sqrt{1\over3}\ktt+\sqrt{2\over3}\kss\\ \nonumber
|8'8'\rangle_c&=&\sqrt{2\over3}\ktt+\sqrt{1\over3}\kss \, .
\end{eqnarray}
\begin{table}[h!]
\caption{Color matrix elements.}
\label{coma}
\begin{center}
\begin{tabular}{|c|cccccc|}
\hline
$\hat{O}$ &$(\vec\lambda_1\cdot\vec\lambda_2)$&$(\vec\lambda_3\cdot\vec\lambda_4)$&$(\vec\lambda_1\cdot\vec\lambda_3)$
&$(\vec\lambda_2\cdot\vec\lambda_4)$&$(\vec\lambda_1\cdot\vec\lambda_4)$&$(\vec\lambda_2\cdot\vec\lambda_3)$\\
\hline
$\btt \hat{O}\ktt$&$-8/3$&$-8/3$&$-4/3$&$-4/3$&$-4/3$&$-4/3$\\
$\bss \hat{O}\kss$&$4/3$&$4/3$&$-10/3$&$-10/3$&$-10/3$&$-10/3$\\
$\btt \hat{O}\kss$&0&0&$-2\sqrt{2}$&$-2\sqrt{2}$&$2\sqrt{2}$&$2\sqrt{2}$\\
\hline
\end{tabular}
\end{center}
\end{table}

To evaluate color matrix elements the two-body color
operators are introduced in the same manner as in angular momentum theory,
\begin{equation}
\label{momen}
\vec\lambda_i\cdot \vec\lambda_j={1\over2}\Big(\vec\lambda^2_{ij}-\vec\lambda_i^2-\vec\lambda_j^2\Big)\, ,
\end{equation}
where $\vec\lambda_i$ are the $SU(3)_c$ Gell-Mann matrices acting
on quark $i$, and $\vec\lambda_{ij}^2$ is the Casimir operator. For an
irreducible representation $\psi(\lambda\mu)$, the eigenvalue of the
Casimir operator is given by:
\begin{equation}
\label{eige}
\vec\lambda_{ij}^2\psi(\lambda\mu)={4\over3}\Big(\lambda^2+\mu^2+\lambda\mu+3\lambda+3\mu\Big)\psi(\lambda\mu)\,.
\end{equation}
In the color space a quark is described by $3_c=(10)$ and an antiquark by $\bar
3_c=(01)$, so
\begin{eqnarray}
\label{ant}
\vec\lambda_i^2\psi(10)&=&\vec\lambda_i^2[3_c]={16\over3}[3_c]={16\over3}\psi(10)\, , \\ \nonumber
\vec\lambda_i^2\psi(01)&=&\vec\lambda_i^2[\bar 3_c]={16\over3}[\bar 3_c]={16\over3}\psi(01)\, .
\end{eqnarray}
Two quarks in a symmetric state, $6$ or $\bar 6$, have $(\lambda\mu)=(20)$ and therefore
\begin{equation}
\label{ant2}
\vec\lambda_i^2\psi(20)=\vec\lambda_i^2[6_c]=\vec\lambda_i^2[\bar 6_c]={40\over3}\psi(20),
\end{equation}
while two quarks in an antisymmetric state, $3$ or $\bar 3$, have
$(\lambda\mu)=(01)$, being the same value as \linebreak Equation
\eref{ant}. Using these expressions, the color matrix elements
summarized in Table~\ref*{coma}, may be easily evaluated.

\subsection{Spin space}
\label{subspin}
The spin part of the wave function can be written as
\begin{equation}
\left[(s_1s_2)_{S_{12}}(s_3s_4)_{S_{34}}\right]_{S}\equiv|S_{12}S_{34}\rangle^{12}_s
\end{equation}
where the spin of the two quarks (antiquarks) is coupled to $S_{12}$ ($S_{34}$).
Two identical spin-$1/2$ fermions in a $S=0$ state are antisymmetric $(A)$ under permutations while those coupled to
$S=1$ are symmetric $(S)$.
In Table~\ref*{spin} we have included the corresponding vectors for each total
spin together with their symmetry properties.
\begin{table}[h!]
\caption[Spin basis vectors.]{Spin basis vectors for all possible
total spin states $(S)$. The ``Symmetry'' column stands for the
symmetry properties of the pair of quarks and antiquarks.}
\label{spin}
\begin{center}
\begin{tabular}{|c|cc|}
\hline
$S$&Vector&Symmetry\\
\hline
0&$|00\rangle^{12}_s$&AA\\
 &$|11\rangle^{12}_s$&SS\\
\hline
 &$|01\rangle^{12}_s$&AS\\
1&$|10\rangle^{12}_s$&SA\\
 &$|11\rangle^{12}_s$&SS\\
\hline
2&$|11\rangle^{12}_s$&SS\\
\hline
\end{tabular}
\end{center}
\end{table}

Using this notation is straightforward to evaluate the four-body
spin matrix elements,
\begin{equation}
_s^{12}\langle S_{12}S_{34}|\vec\sigma_i\cdot\vec\sigma_j|S_{12}'S_{34}'\rangle_s^{12}=
\Big[2S_{ij}(S_{ij}+1)-3\Big]\delta_{S_{12},S_{12}'}\delta_{S_{34},S_{34}'}\delta_{S,S'} \, ,
\end{equation}
for $(ij)=(12)$ or (34) and where $\vec\sigma_i$ is the spin operator acting over particle $i$.
To calculate the other spin operators we should reorder the spin wave function~\cite{Va88}
\begin{eqnarray}
\Big[(s_1s_2)_{S_{12}}(s_3s_4)_{S_{34}}\Big]_{S}
&=&\sum_{k,l}(-1)^{2S_{12}+s_2+2s_3+s_4+l+S}\sqrt{2k+1}\sqrt{2l+1}\sqrt{2S_{12}+1}\sqrt{2S_{34}+1} \nonumber \\
&&\left\{\begin{array}{ccc}S_{12}&s_3&k\\s_4&S&S_{34}\end{array}\right\}
\left\{\begin{array}{ccc}s_2&s_1&S_{12}\\s_3&k&l\end{array}\right\}\Bigg[\Big[(s_1s_3)_ls_2\Big]_ks_4\Bigg]_{S}.
\end{eqnarray}
Now one can calculate the matrix element for the case $s_1=s_2=s_3=s_4={1\over2}$,
\begin{eqnarray}
&&_s^{12}\langle S_{12}S_{34}|\vec\sigma_1\cdot\vec\sigma_3|S_{12}'S_{34}'\rangle_s^{12}=\\ \nonumber
&=&\sqrt{2S_{12}+1}\sqrt{2S'_{12}+1}\sqrt{2S_{34}+1}\sqrt{2S'_{34}+1}\sum_{k,\,l}(2k+1)(2l+1)\big[2l(l+1)-3\big]\times\\ \nonumber
&\times&\left\{\begin{array}{ccc}S_{12}&1/2&k\\1/2&S&S_{34}\end{array}\right\}
\left\{\begin{array}{ccc}S'_{12}&1/2&k\\1/2&S&S'_{34}\end{array}\right\}
\left\{\begin{array}{ccc}1/2&1/2&S_{12}\\1/2&k&l\end{array}\right\}
\left\{\begin{array}{ccc}1/2&1/2&S'_{12}\\1/2&k&l\end{array}\right\}.
\end{eqnarray}
The same can be done for the other spin operators, $(\vec\sigma_1\cdot\vec\sigma_4)$, $(\vec\sigma_2\cdot\vec\sigma_4)$
and $(\vec\sigma_2\cdot\vec\sigma_3)$, using the expressions given above. The results are resumed in Table~\ref*{tabs}.
\begin{table}[h!!]
\caption{Spin matrix elements.}
\label{tabs}
\begin{center}
\begin{tabular}{|c|c|cccccc|}
\hline
$S$&&$(\vec\sigma_1\cdot\vec\sigma_2)$&$(\vec\sigma_3\cdot\vec\sigma_4)$&$(\vec\sigma_1\cdot\vec\sigma_3)$
&$(\vec\sigma_2\cdot\vec\sigma_4)$ &$(\vec\sigma_1\cdot\vec\sigma_4)$&$(\vec\sigma_2\cdot\vec\sigma_3)$\\
\hline
&$^{12}_s\langle 00|\hat{O}|00\rangle^{12}_s$&$-3$&$-3$&0&0&0&0\\
0&$^{12}_s\langle 11|\hat{O}|11\rangle^{12}_s$&$1$&$1$&$-2$&$-2$&$-2$&$-2$\\
&$^{12}_s\langle 00|\hat{O}|11\rangle^{12}_s$&0&0&$-\sqrt{3}$&$-\sqrt{3}$&$\sqrt{3}$&$\sqrt{3}$\\
\hline
&$^{12}_s\langle 01|\hat{O}|01\rangle^{12}_s$&$-3$&$1$&0&0&0&0\\
&$^{12}_s\langle 10|\hat{O}|10\rangle^{12}_s$&$1$&$-3$&0&0&0&0\\
1&$^{12}_s\langle 11|\hat{O}|11\rangle^{12}_s$&$1$&$1$&$-1$&$-1$&$-1$&$-1$\\
&$^{12}_s\langle 01|\hat{O}|10\rangle^{12}_s$&0&0&1&1&$-1$&$-1$\\
&$^{12}_s\langle 10|\hat{O}|11\rangle^{12}_s$&0&0&$\sqrt{2}$&$-\sqrt{2}$&$-\sqrt{2}$&$\sqrt{2}$\\
&$^{12}_s\langle 01|\hat{O}|11\rangle^{12}_s$&0&0&$-\sqrt{2}$&$\sqrt{2}$&$-\sqrt{2}$&$\sqrt{2}$\\
\hline
2&$^{12}_s\langle 11|\hat{O}|11\rangle^{12}_s$&1&1&1&1&1&1\\
\hline
\end{tabular}
\end{center}
\end{table}

\begin{table}[h!!]
\caption{Pauli-based classification of four-quark states.
$\checkmark$ indicates that the quark/antiquark pair requires the
application of the Pauli principle, being the notation (pair of
quarks, pair of antiquarks). The third and fourth columns contain
the recoupling corresponding to bases~\eref{eq1b} and~\eref{eq1c}.}
\label{tab_clasi}
\begin{center}
\begin{tabular}{|c|c||cc|}
\hline
(12)(34)            & Pauli             &(13)(24)           & (14)(23) \\
\hline
$(nn)(\bar n\bar n)$        & $(\checkmark,\checkmark)$     & $(n\bar n)(n\bar n)$      & $(n\bar n)(n\bar n)$\\
$(nn)(\bar n\bar Q)$        & $(\checkmark,X)$      & $(n\bar n)(n\bar Q)$      & $(n\bar Q)(n\bar n)$\\
$(nn)(\bar Q_1\bar Q_2)$    & $(\checkmark,\checkmark$ if $\bar Q_1=\bar Q_2$)
                                & $(n\bar Q_1)(n\bar Q_2)$  & $(n\bar Q_2)(n\bar Q_1)$\\
$(nQ_1)(\bar n\bar Q_2)$    & $(X,X$)           & $(n\bar n)(Q_1\bar Q_2)$  & $(n\bar Q_2)(Q_1\bar n)$\\
$(nQ_1)(\bar Q_2\bar Q_3)$  & $(X,\checkmark$ if $\bar Q_2=\bar Q_3$)
                                & $(n\bar Q_2)(Q_1\bar Q_3)$    & $(n\bar Q_3)(Q_1\bar Q_2)$\\
$(Q_1Q_2)(\bar Q_3\bar Q_4)$    & $(\checkmark$ if $Q_1=Q_2$, $\checkmark$ if $\bar Q_3=\bar Q_4$)
                                & $(Q_1\bar Q_3)(Q_2\bar Q_4)$  & $(Q_1\bar Q_4)(Q_2\bar Q_3)$\\
\hline
\end{tabular}
\end{center}
\end{table}

\subsection{Flavor space}
\label{flwave}

Before discussing the flavor part of the wave function one must
specify the required flavor symmetry, $SU(2)$ or $SU(3)$.
In the former case, $u$ and $d$ quarks are identical
whether in the latter, $u$, $d$, and $s$ are indistinguishable.
In the following, $n$ will stand for light $u$ and $d$ quarks and $Q$
for heavy ones, $c$ or $b$. $s$ quarks will be considered {\em heavy}
if flavor $SU(2)$ is assumed and {\em light} otherwise.

For the flavor part one finds several different possible four-quark
states depending on the number of light quarks. They can be
classified depending on whether they are made of undistinguishable
quarks in one of the pairs (and therefore the Pauli principle must
be imposed) or not. In following subsections we will discuss the
important role played by the Pauli principle in the description of
the four-quark states properties. This classification is illustrated
in Table~\ref*{tab_clasi}. Symmetry properties of the flavor wave
function are summarized in Table~\ref*{sim_flav}.
\begin{table}[h!!]
\caption{Symmetry properties of the flavor wave function under the
exchange of quarks (the same holds for antiquarks). $^{\dagger}$ If
flavor $SU(3)$ is assumed, symmetric and antisymmetric flavor wave
functions with $I=1/2$ can be constructed, i.e., $(us\pm
su)/\sqrt{2}$).} \label{sim_flav}
\begin{center}
\begin{tabular}{|c|c|}
\hline
Flavor&Symmetry\\
\hline
$nn$ $I=0$      & A \\
$nn$ $I=1$      & S \\
$nn$ $I=1/2^{\dagger}$  & S/A   \\
$QQ$ $I=0$      & S \\
\hline
\end{tabular}
\end{center}
\end{table}

The flavor $SU(2)$ matrix elements can be evaluated by means of the
same relations shown in Section~\ref*{subspin}. For those
corresponding to flavor $SU(3)$ the procedure will require the
explicit construction of the flavor wave function by means of the
$SU(3)$ isoscalar factors given in~\cite{De63a,De63b}\footnote{Note there
is no universal agreement in the phase convention regarding the
isoscalar factor, so mixing different tables from different authors
should be done with care.}. As an example we evaluate some of the
flavor matrix elements needed for the description of heavy-light
tetraquarks. They can be obtained using the matrix expression of
$\lambda^a$,
\begin{eqnarray}
&&\lambda^1=\left(\begin{array}{ccc}0&1&0\\1&0&0\\0&0&0\end{array}\right)\,\,\,\,\,\,\,\,
\lambda^2=\left(\begin{array}{ccc}0&-i&0\\i&0&0\\0&0&0\end{array}\right)\,\,\,\,\,\,\,\,
\lambda^3=\left(\begin{array}{ccc}1&0&0\\0&-1&0\\0&0&0\end{array}\right)\\ \nonumber
&&\lambda^4=\left(\begin{array}{ccc}0&0&1\\0&0&0\\1&0&0\end{array}\right)\,\,\,\,\,\,\,\,
\lambda^5=\left(\begin{array}{ccc}0&0&-i\\0&0&0\\i&0&0\end{array}\right)\,\,\,\,\,\,\,\,
\lambda^6=\left(\begin{array}{ccc}0&0&0\\0&0&1\\0&1&0\end{array}\right)\\ \nonumber
&&\lambda^7=\left(\begin{array}{ccc}0&0&0\\0&0&-i\\0&i&0\end{array}\right)\,\,\,\,\,\,\,\,
\lambda^8=\left(\begin{array}{ccc}{1\over\sqrt{3}}&0&0\\0&{1\over\sqrt{3}}&0\\0&0&{-2\over\sqrt{3}}\end{array}\right),
\end{eqnarray}
where, following the same convention, quarks and antiquarks are given by,
\begin{eqnarray}
&u&=\bar u=(1,0,0)\\ \nonumber
&d&=\bar d=(0,1,0)\\ \nonumber
&s&=\bar s=(0,0,1).
\end{eqnarray}
The tetraquark flavor wave function corresponding to two light quarks coupled to total isospin
$I$ with $I_z=0$ and two heavy antiquarks can be written as
\begin{equation}
\left|\psi\et={1\over\sqrt{2}}[ud+(-1)^{I+1}du][\bar s\bar s]\,.
\end{equation}
A typical flavor operator is
\begin{equation}
\label{opeI}
\vec\tau_i\cdot\vec\tau_j=\sum_{a=1}^3\lambda_i^a\lambda_j^a\, ,
\end{equation}
where $\lambda_i^a$ are the $SU(3)$ flavor matrices defined above and $\tau_i$ are the isospin
Pauli matrices, both acting on quark $i$. So the same expression obtained
for the spin operators holds here:
\begin{equation}
\br \psi\Big|\sum_{a=1}^3\lambda^a_1\lambda^a_2\Big|\psi\et=
\left\{\begin{array}{ll}I=0&\rightarrow-3\\I=1&\rightarrow1\end{array}\right.\, .
\end{equation}
Alternatively one can write the flavor matrix element as
\begin{eqnarray}
\label{flqq}
\br \psi|\sum_{a=1}^3\lambda^a_1\lambda^a_2|\psi\et
&=&\br {ud+(-1)^{I+1}du\over\sqrt{2}}\Big|\sum_{a=1}^3\lambda^a_1\lambda^a_2\Big|{ud+(-1)^{I+1}du\over\sqrt{2}}\et=\\ \nonumber
&=&{1\over2}\sum_{a=1}^3\Bigg\{\br ud\Big|\lambda^a_1\lambda^a_2\Big|ud\et+\br
du\Big|\lambda^a_1\lambda^a_2\Big|du\et+\\ \nonumber
&+&(-1)^{I+1}\br du\Big|\lambda^a_1\lambda^a_2\Big|ud\et+(-1)^{I+1}\br
ud\Big|\lambda^a_1\lambda^a_2\Big|du\et\Bigg\}=\\ \nonumber
&=&\sum_{a=1}^3\Bigg\{\br u|\lambda^a|u\et\br d|\lambda^a|d\et+(-1)^{I+1}|\br u|\lambda^a|d\et|^2\Bigg\}=\\ \nonumber
&=&-1+2(-1)^{I+1}=\left\{\begin{array}{ll}I=0&-3\\I=1&1\end{array}\right. \, .
\end{eqnarray}
Other matrix elements of interest are,
\begin{eqnarray}
&&\br \psi|\lambda_1^8\lambda_2^8| \psi\et={1\over 3}\\ \nonumber
&&\br \psi|\lambda_3^8\lambda_4^8| \psi\et={4\over 3}\\ \nonumber
&&\br \psi|\lambda_1^8\lambda_3^8| \psi\et=
\br \psi|\lambda_2^8\lambda_3^8| \psi\et=
\br \psi|\lambda_1^8\lambda_4^8| \psi\et=
\br \psi|\lambda_2^8\lambda_4^8| \psi\et=-{2\over 3}\, .
\end{eqnarray}

\subsection{Radial space}

The most general radial wave function with orbital angular momentum
$L=0$ may depend on the six scalar quantities that can be
constructed with the Jacobi coordinates of the system, they are:
$\vec x^{\,2}$, $\vec y^{\,2}$, $\vec z^{\,2}$,
$\vec{x}\cdot\vec{y}$, $\vec{x}\cdot\vec{z}$ and
$\vec{y}\cdot\vec{z}$. We define the variational spatial wave
function as a linear combination of \linebreak {\em generalized
Gaussians},
\begin{equation}
\left|R_{s_4}\right>=\sum_{i=1}^{n} \beta_{s_4}^{(i)} R_{s_4}^i(\vec x,\vec y,\vec z)=\sum_{i=1}^{n} \beta_{s_4}^{(i)} R_{s_4}^i
\label{wave}
\end{equation}
where $n$ is the number of Gaussians we use for each
color-spin-flavor component. $R_{s_4}^i$ depends on six variational
parameters, $a^i_s$, $b^i_s$, $c^i_s$, $d^i_s$, $e^i_s$, and
$f^i_s$, one for each scalar quantity. Therefore, any tetraquark
\pagebreak will depend on $6\times n\times n_s$ variational
parameters (where $n_s$ is the number of different channels allowed
by the Pauli Principle). Equation~\eref{wave} should have well
defined permutation symmetry under the exchange of both quarks and
antiquarks,
\begin{eqnarray}
\label{parx}
P_{12}(\vec x   \rightarrow -\vec x)R^i_{s_4}&=&P_xR^i_{s_4}\\ \nonumber
P_{34}(\vec y   \rightarrow -\vec y)R^i_{s_4}&=&P_yR^i_{s_4},
\end{eqnarray}
where $P_x$ and $P_y$ are $-1$ for antisymmetric states, $(A)$, and $+1$ for symmetric ones, $(S)$. One can build
the following radial combinations, $(P_xP_y)=(SS)$, $(SA)$, $(AS)$ and $(AA)$:
\begin{eqnarray}
\label{wave2-1}
(SS)\Rightarrow R_1^i&=&
{\rm Exp}\left(-a^i_s\vec x^{\,2}-b^i_s\vec y^{\,2}-c^i_s\vec z^{\,2}-d^i_s\vec x\vec y-e^i_s\vec x\vec z-f^i_s\vec y\vec z\right)\\\nonumber
&+&{\rm Exp}\left(-a^i_s\vec x^{\,2}-b^i_s\vec y^{\,2}-c^i_s\vec z^{\,2}+d^i_s\vec x\vec y-e^i_s\vec x\vec z+f^i_s\vec y\vec z\right)\\\nonumber
&+&{\rm Exp}\left(-a^i_s\vec x^{\,2}-b^i_s\vec y^{\,2}-c^i_s\vec z^{\,2}+d^i_s\vec x\vec y+e^i_s\vec x\vec z-f^i_s\vec y\vec z\right)\\\nonumber
&+&{\rm Exp}\left(-a^i_s\vec x^{\,2}-b^i_s\vec y^{\,2}-c^i_s\vec z^{\,2}-d^i_s\vec x\vec y+e^i_s\vec x\vec z+f^i_s\vec y\vec z\right) \, ,
\end{eqnarray}
\begin{eqnarray}
\label{wave2-2}
(SA)\Rightarrow R_2^i&=&
{\rm Exp}\left(-a^i_s\vec x^{\,2}-b^i_s\vec y^{\,2}-c^i_s\vec z^{\,2}-d^i_s\vec x\vec y-e^i_s\vec x\vec z-f^i_s\vec y\vec z\right)\\\nonumber
&-&{\rm Exp}\left(-a^i_s\vec x^{\,2}-b^i_s\vec y^{\,2}-c^i_s\vec z^{\,2}+d^i_s\vec x\vec y-e^i_s\vec x\vec z+f^i_s\vec y\vec z\right)\\\nonumber
&+&{\rm Exp}\left(-a^i_s\vec x^{\,2}-b^i_s\vec y^{\,2}-c^i_s\vec z^{\,2}+d^i_s\vec x\vec y+e^i_s\vec x\vec z-f^i_s\vec y\vec z\right)\\\nonumber
&-&{\rm Exp}\left(-a^i_s\vec x^{\,2}-b^i_s\vec y^{\,2}-c^i_s\vec z^{\,2}-d^i_s\vec x\vec y+e^i_s\vec x\vec z+f^i_s\vec y\vec z\right) \, ,
\end{eqnarray}
\begin{eqnarray}
\label{wave2-3}
(AS)\Rightarrow R_3^i&=&
{\rm Exp}\left(-a^i_s\vec x^{\,2}-b^i_s\vec y^{\,2}-c^i_s\vec z^{\,2}-d^i_s\vec x\vec y-e^i_s\vec x\vec z-f^i_s\vec y\vec z\right)\\\nonumber
&+&{\rm Exp}\left(-a^i_s\vec x^{\,2}-b^i_s\vec y^{\,2}-c^i_s\vec z^{\,2}+d^i_s\vec x\vec y-e^i_s\vec x\vec z+f^i_s\vec y\vec z\right)\\\nonumber
&-&{\rm Exp}\left(-a^i_s\vec x^{\,2}-b^i_s\vec y^{\,2}-c^i_s\vec z^{\,2}+d^i_s\vec x\vec y+e^i_s\vec x\vec z-f^i_s\vec y\vec z\right)\\\nonumber
&-&{\rm Exp}\left(-a^i_s\vec x^{\,2}-b^i_s\vec y^{\,2}-c^i_s\vec z^{\,2}-d^i_s\vec x\vec y+e^i_s\vec x\vec z+f^i_s\vec y\vec z\right) \, ,
\end{eqnarray}
\begin{eqnarray}
\label{wave2-4}
(AA)\Rightarrow R_4^i&=&
{\rm Exp}\left(-a^i_s\vec x^{\,2}-b^i_s\vec y^{\,2}-c^i_s\vec z^{\,2}-d^i_s\vec x\vec y-e^i_s\vec x\vec z-f^i_s\vec y\vec z\right)\\\nonumber
&-&{\rm Exp}\left(-a^i_s\vec x^{\,2}-b^i_s\vec y^{\,2}-c^i_s\vec z^{\,2}+d^i_s\vec x\vec y-e^i_s\vec x\vec z+f^i_s\vec y\vec z\right)\\\nonumber
&-&{\rm Exp}\left(-a^i_s\vec x^{\,2}-b^i_s\vec y^{\,2}-c^i_s\vec z^{\,2}+d^i_s\vec x\vec y+e^i_s\vec x\vec z-f^i_s\vec y\vec z\right)\\\nonumber
&+&{\rm Exp}\left(-a^i_s\vec x^{\,2}-b^i_s\vec y^{\,2}-c^i_s\vec z^{\,2}-d^i_s\vec x\vec y+e^i_s\vec x\vec z+f^i_s\vec y\vec z\right) \, .
\end{eqnarray}
By defining the function
\begin{equation}
\label{red1}
g(s_1,s_2,s_3)={\rm Exp}\left(-a^i_s\vec x^{\,2}-b^i_s\vec y^{\,2}-c^i_s\vec z^{\,2}
-s_1d^i_s\vec x\vec y-s_2e^i_s\vec x\vec z-s_3f^i_s\vec y\vec z\right),
\end{equation}
we can build the vectors
\begin{equation}
\vec{G_s^i}=\left(\begin{array}{l} g(+,+,+)\\g(-,+,-)\\g(-,-,+)\\g(+,-,-)\end{array}\right)\, ,
\end{equation}
and
\begin{eqnarray}
\label{red2}
\vec{\alpha_{SS}}&=&(+,+,+,+)\\ \nonumber
\vec{\alpha_{SA}}&=&(+,-,+,-)\\ \nonumber
\vec{\alpha_{AS}}&=&(+,+,-,-)\\ \nonumber
\vec{\alpha_{AA}}&=&(+,-,-,+),
\end{eqnarray}
what allows to write in a compact way
Equations~\eref{wave2-1},~\eref{wave2-2},~\eref{wave2-3},
and~\eref{wave2-4},
\begin{eqnarray}
\label{redu}
(SS)&\Rightarrow& R_1^i=\vec{\alpha_{SS}}\cdot\vec{G_s^i}\\ \nonumber
(SA)&\Rightarrow& R_2^i=\vec{\alpha_{SA}}\cdot\vec{G_s^i}\\ \nonumber
(AS)&\Rightarrow& R_3^i=\vec{\alpha_{AS}}\cdot\vec{G_s^i}\\ \nonumber
(AA)&\Rightarrow& R_4^i=\vec{\alpha_{AA}}\cdot\vec{G_s^i} \, .
\end{eqnarray}
Such a radial wave function includes all possible relative orbital angular momenta
coupled to $L=0$. This can be seen through the relation:
\begin{eqnarray}
\label{ldz}
{\rm Exp}\left(-d^i_s\vec x\vec y-e^i_s\vec x\vec z-f^i_s\vec y\vec
z\right)={1\over\sqrt{4\pi}}\sum_{\ell_x=0}^\infty\sum_{\ell_y=0}^\infty\sum_{\ell_z=0}^\infty
\left[[Y_{\ell_x}(\hat x)Y_{\ell_y}(\hat y)]_{\ell_z}Y_{\ell_z}(\hat
z)\right]_0\\ \nonumber
\sum_{\ell_1,\ell_2,\ell_3}(2\ell_1+1)(2\ell_2+1)(2\ell_3+1)\br \ell_10\ell_20|\ell_x\et\br
\ell_10\ell_30|\ell_y\et\br
\ell_20\ell_30|\ell_z\et\left\{\begin{array}{ccc}\ell_x&\ell_y&\ell_z\\\ell_3&\ell_2&\ell_1\end{array}\right\}
\\ \nonumber
\left(\sqrt{\pi\over {2\,d_s^ixy}}I_{\ell_1+1/2}(d_s^ixy)\right)
\left(\sqrt{\pi\over {2\,e_s^ixz}}I_{\ell_2+1/2}(e_s^ixz)\right)
\left(\sqrt{\pi\over {2\,f_s^iyz}}I_{\ell_3+1/2}(f_s^iyz)\right) \, ,
\end{eqnarray}
where $\ell_x$, $\ell_y$ and $\ell_z$ are the orbital angular momenta associated
to coordinates $\vec x$, $\vec y$ and $\vec z$, and $I_a(x)$ are the modified Bessel functions.

The radial wave functions defined above have also well-defined
symmetry properties on the $\vec z$ coordinate. Being
$P_{(12)(34)}(\vec z \rightarrow -\vec z)R^i_{s_4}=P_zR^i_{s_4}$
one obtains,
\begin{eqnarray}
 \label{parz}
 P_{(12)(34)}R_1^i&=&+R_1^i\\ \nonumber
 P_{(12)(34)}R_2^i&=&-R_2^i\\ \nonumber
 P_{(12)(34)}R_3^i&=&-R_3^i\\ \nonumber
 P_{(12)(34)}R_4^i&=&+R_4^i \, .
\end{eqnarray}
To evaluate radial matrix elements we will use the notation
introduced in Equation~\eref{redu}:
\begin{equation}
\label{ra1}
\br R_{\gamma}^i|f(x,y,z)|R_{\beta}^j\et=\int_V(\vec \alpha_{S_\gamma}\cdot \vec G^i_s)f(x,y,z)(\vec \alpha_{S_\beta}\cdot \vec G^j_{s'})dV=
\vec \alpha_{S_\gamma}\cdot F^{ij}\cdot\vec \alpha_{S_\beta}\, ,
\end{equation}
where $\gamma$ and $\beta$ stand for the symmetry of the radial
wave function and $F^{ij}$ is a matrix whose element $(a,b)$ is
defined through,
\begin{equation}
F^{ij}_{ab}=\int_V(\vec G_s^i)_a(\vec G^j_{s'})_bf(x,y,z)dV\, ,
\end{equation}
being $(\vec G_s^i)_a$ the component $a$ of the vector $\vec G_s^i$.
>From Equation~\eref{red1} one obtains,
\begin{equation}
g(s_1,s_2,s_3)g(s'_1,s'_2,s'_3)={\rm Exp}\left(-a_{ij}\vec x^{\,2}-b_{ij}\vec y^{\,2}-c_{ij}\vec z^{\,2}
-\bar s_{ij}\vec x\vec y-\bar e_{ij}\vec x\vec z-\bar f_{ij}\vec y\vec z\right) \, ,
\end{equation}
where we have shortened the previous notation according to $a^i_s\to a_i$,
$a_{ij}=a_i+a_j$ and $\bar d_{ij}=(s_1d_i+s_1'd_j)$. Therefore,
all four-body radial matrix elements will contain integrals of the form
\begin{equation}
I=\int_V{\rm Exp}\left(-a_{ij}\vec x^{\,2}-b_{ij}\vec y^{\,2}-c_{ij}\vec
z^{\,2} -\bar s_{ij}\vec x\vec y-\bar e_{ij}\vec x\vec z-\bar f_{ij}\vec y\vec z\right)f(x,y,z)d\vec xd\vec yd\vec z  \, ,
\end{equation}
where the functions $f(x,y,z)$ are the potentials. Being
all of them radial functions (not depending on angular variables)
one can solve the previous integral by noting:
\begin{equation}
\int {\rm Exp}\big[-\sum_{i,j=1}^nA_{ij}\vec x_i \vec x_j\big]f\big(|\sum \alpha_k\vec
x_k|\big)d\vec x_1...d\vec
x_n=\Bigg({\pi^n\over{det\,A}}\Bigg)^{3\over2}4\pi\Bigg({\Omega_{ij}\over\pi}\Bigg)^{3\over2}F(\Omega_{ij},f) \, ,
\end{equation}
where
\begin{eqnarray}
{1\over\Omega_{ij}}&=&\bar\alpha\cdot A^{-1} \cdot\alpha\\ \nonumber
F(A,f)&=&\int e^{-Au^2}f(u)u^2du\\ \nonumber
det\,A&>&0\\ \nonumber
{1\over\Omega_{ij}}&>&0 \, .
\end{eqnarray}
One can extract some useful relations for the radial matrix elements
using simple symmetry properties. Let us rewrite Equation~\eref{ra1}
\begin{eqnarray}
\br R_{\gamma}^i|f(x,y,z)|R_{\beta}^j\et&=&\br R_{P_xP_yP_z}^i|f(x,y,z)|R_{P_x'P_y'P_z'}^j\et\\ \nonumber
&=&\int_x\int_y\int_z R_{P_xP_yP_z}^if(x,y,z)R_{P_x'P_y'P_z'}^jd\vec xd\vec yd\vec z \, .
\end{eqnarray}
If $f(x,y,x)$ depends only in one coordinate, for example $\vec x$, the
integrals over the other coordinates will be zero if one of them has different symmetry properties,
$P_y\neq P_y'$ or $P_z\neq P_z'$ in our example. Therefore
\begin{eqnarray}
\br R_{\gamma}^i|f(x)|R_{\beta}^j\et&\propto&\delta_{\gamma\beta}\\ \nonumber
\br R_{\gamma}^i|f(y)|R_{\beta}^j\et&\propto&\delta_{\gamma\beta}\\ \nonumber
\br R_{\gamma}^i|f(z)|R_{\beta}^j\et&\propto&\delta_{\gamma\beta}\\ \nonumber
\br R_{\gamma}^i|{\rm Constant}|R_{\beta}^j\et&\propto&\delta_{\gamma\beta} \, .
\end{eqnarray}
The radial wave function described in this section is adequate to describe
not only bound states, but also it is flexible enough to describe states of
the two-meson continuum within a reasonable accuracy.
We will came back to this point in Sect.~\ref*{results}

\subsection{Parity and $C-$parity}

The parity of a tetraquark can be calculated as
\begin{equation}
\label{pari1}
P\left[R_{s_4}^i(\vec x,\vec y,\vec z)\right]=R_{s_4}^i\left(\begin{array}{c}\vec x\to-\vec x\\\vec y\to-\vec y\\
\vec z\to-\vec z\end{array}\right)=(-1)^{\ell_x+\ell_y+\ell_z}R_{s_4}^i(\vec x,\vec y,\vec z)\, ,
\end{equation}
or using Equations~\eref{parx} and~\eref{parz},
\begin{equation}
\label{pari2b}
P\left[R_{s_4}^i(\vec x,\vec y,\vec z)\right]=P_{12}P_{34}P_{(12)(34)}R_{s_4}^i(\vec x,\vec y,\vec z)=P_xP_yP_z\,
R_{s_4}^i(\vec x,\vec y,\vec z)\, ,
\end{equation}
what in our case implies
\begin{equation}
P\left[R_{s_4}^i(\vec x,\vec y,\vec z)\right] =
\left\{\begin{array}{c} (+)(+)(+)R^i_1 \\
                        (+)(-)(-)R^i_2 \\
                        (-)(+)(-)R^i_3 \\
                        (-)(-)(+)R^i_4
\end{array}\right\}=
\left\{\begin{array}{c} +R^i_1 \\
                        +R^i_2 \\
                        +R^i_3 \\
                        +R^i_4
\end{array}\right\}=
+R_{s_4}^i(\vec x,\vec y,\vec z) \, .
\label{pari2}
\end{equation}
Hence, this formalism describes positive parity states, being thus
adequate to study tetraquark ground states.
\begin{table}[h!!]
\caption{The action of $C$ over the spin part or the wave function.}
\label{c_spin}
\begin{center}
\begin{tabular}{|cc|}
\hline
$S=0$   &$C|00\rangle^{12}_s=+|00\rangle^{12}_s$\\
    &$C|11\rangle^{12}_s=+|11\rangle^{12}_s$\\
\hline
    &$C|01\rangle^{12}_s=-|10\rangle^{12}_s$\\
$S=1$   &$C|10\rangle^{12}_s=-|01\rangle^{12}_s$\\
    &$C|11\rangle^{12}_s=+|11\rangle^{12}_s$\\
\hline
$S=2$   &$C|11\rangle^{12}_s=+|11\rangle^{12}_s$\\
\hline
\end{tabular}
\end{center}
\end{table}

From Equation~\eref{pari2} one can see that, not only the total wave
function will be an eigenstate of the parity operator, but also each
component will be. This is not the case for $C$-parity, where only
the total wave function will be an eigenstate and therefore it must
be obtained numerically for each state. The tetraquark $C-$parity
will depend on the variational parameters and on the
$\beta_{s}^{(i)}$ coefficients. This dependence is contained in the
action of the $C-$parity operator over the different parts of the
wave function which will give us the following relations:
\begin{equation}
C\left|R_{s_4}^i(\vec x,\vec y,\vec z)\et=R_{s_4}^i(\vec y,\vec x,-\vec z) \, ,
\end{equation}
and if $a_s^i=b_s^i$ and $e_s^i=f_s^i$ (what is a very common result),
\begin{equation}
C\left|R_{s_4}^i(\vec x,\vec y,\vec z)\et=\left\{\begin{array}{cc}s_4=1&\rightarrow +R_1^i\\
                        s_4=2&\rightarrow -R_3^i\\
                        s_4=3&\rightarrow -R_2^i\\
                        s_4=4&\rightarrow +R_4^i\end{array}\right..
\end{equation}
The action of $C$ over the spin part or the wave function is summarized in Table~\ref*{c_spin}.
The action of $C$ over the flavor part of the wave function has to be evaluated individually once the wave
functions have been constructed.

\subsection{$q\bar q\leftrightarrow qq\bar q\bar q$ mixing}
\label{mixing}

Many of the possible four-quark systems may present $J^{PC}$ quantum numbers that can be
reached not only by means of $qq\bq\bq$ configurations but also by $q\bq$ ones and with similar energies.
In these cases the possibility of a mixing between them cannot be discarded a priori and therefore, their most
general wave function will read
\begin{equation}
\label{mes-w}
\left|\mathrm{B=0}\et=\sum_n\Omega_n\left|(q\bq)^n\et=\Omega_1\left|q\bq\et+\Omega_2\left|q\bq q\bq\et+....
\end{equation}
These particular systems may be described by a hamiltonian
\begin{equation}
H=H_0 + H_1 \,\,\,\,\, {\rm being} \,\,\,\,\,
H_0=\left(\begin{array}{cc}H_{q\bq} & 0 \\ 0 & H_{qq\bq\bq}\end{array}\right)\,\,\,
H_1=\left(\begin{array}{cc} 0 & V_{q\bq \leftrightarrow qq\bq\bq} \\ V_{q\bq \leftrightarrow qq\bq\bq} & 0  \end{array}\right)
\label{per1}
\end{equation}
where the nondiagonal terms can be treated perturbatively, therefore allowing to solve the two- and four-body sectors
separately.
\begin{figure}[h!!]
\begin{center}
\caption{Coupling between $qq\bq\bq$ and $q\bq$ configurations.}
\label{figcq}
\epsfig{file=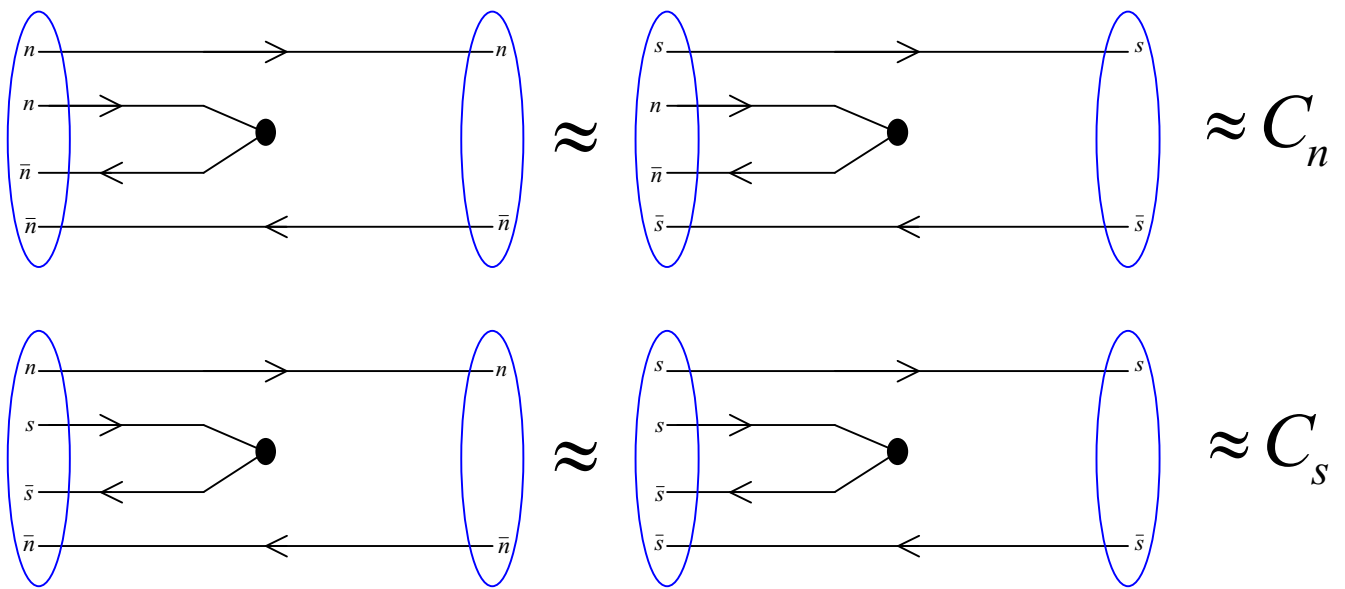}
\end{center}
\vspace*{-0.5cm}
\end{figure}

The Hamiltonian $H_1$ describes the mixing between two- and
four-body configurations. Its explicit expression would require the knowledge
of the operator annihilating a quark-antiquark pair into the vacuum. This could
be done, for example, using a $^3P_0$ model, but the result will introduce an
additional degree of uncertainty on the parametrization used to describe the vertex. Such a parametrization
is determined by the energy scale at which the transition $qq\bq\bq\leftrightarrow q\bq$ takes place.
For the sake of simplicity this can be parametrized by looking to the quark pair that it is annihilated,
and not to the spectator quarks that will form the final $q\bar q$ state:
\begin{equation}
V_{q\bq \leftrightarrow qq\bq\bq}\Rightarrow\left\{
        \begin{array}{l}
        \br nn\bn\bn|V|n\bn\et=\br ns\bn\bs|V|s\bs\et=\br nn\bn\bs|V|n\bs\et=C_n \\
        \br ss\bs\bs|V|s\bs\et=\br ns\bn\bs|V|n\bn\et=\br ns\bs\bs|V|n\bs\et=C_s\end{array}\right.\, .
\end{equation}
A sketch of these mixing interactions is drawn in
Figure~\ref*{figcq}. Such approach has been used in a series of
papers to describe the light-scalar mesons and the open-charm and
open-bottom meson \linebreak sectors~\cite{Vij05,Vij06,Vij08,Vij09}.

\begin{center}
\begin{table}[h!!]
\caption{Lowest two-meson thresholds in the uncoupled (UN) and
coupled (CO) schemes for two particular $cn\bar c\bar n$ (upper) and
$cc\bar n\bar n$ (lower) states, see text for details. They have
been calculated using the CQC model, see Section~\ref*{results} for
details. $M_1\,M_2\vert_L$ indicates the lowest threshold and $L$
its relative orbital angular momentum. Energies are in MeV.}
\label{threstab}
\begin{tabular}{|c||c|c|c|c|c|c||c|c|}
\hline
    & \multicolumn{6}{c||}{UN} & \multicolumn{2}{c|}{CO} \\
\hline
$S$        &0          &1  &2  &0  &1  &2  &\multicolumn{2}{c|}{}\\
\hline
$I$    &\multicolumn{3}{c|}{$0$}    & \multicolumn{3}{c||}{$1$} & 0     & 1\\
\hline
$J^{PC}=1^{++}$ & $-$                   & $J/\psi\,\omega\vert_S$       & $-$   & $-$           & $J/\psi\,\rho\vert_S$ &$-$
        & $J/\psi\,\omega\vert_{S,D}$ & $J/\psi\,\rho\vert_{S,D}$ \\
$(L=0)$     & $-$                   & 3745                  & $-$   & $-$                   & 3838 & $-$
        & 3745                       & 3838\\
\hline
$J^{PC}=1^{--} $& $D\,\bar D\vert_P$            & $\eta_c\,\omega\vert_P$& $D^*\,\bar D^*\vert_P$       & $D\,\bar D\vert_P$
            & $J/\psi\,\pi\vert_P$          & $D^*\,\bar D^*\vert_P$ & $\eta_c\,\omega\vert_P$
        & $J/\psi\,\pi\vert_P$       \\
$(L=1)$     & 3872                  & 3683          & 4002          & 3872
        & 3590          & 4002          & 3683                  & 3590\\
\hline
\hline
$J^P=1^+$ & $-$ & $D\,D^*\vert_S$ & $-$ & $-$ & $D\,D^*\vert_S$ & $-$ & $D\,D^*\vert_{S,D}$ & $D\,D^*\vert_{S,D}$ \\
$(L=0)$   & & 3937 & & & 3937 & & 3937 & 3937 \\
\hline
$J^P=1^-$ & $D\,D\vert_P$ & $D\,D^*\vert_P$ & $D^*\,D^*\vert_P$ & $D\,D_1\vert_S$ & $D\,D^*\vert_P$
      & $D^*\,D_J^*\vert_{S,D}$ & $D\,D\vert_P$ & $D\,D^*\vert_P$ \\
$(L=1)$   & 3872      & 3937        & 4002      & 4426            & 3937
      & 4499            & 3872 & 3937 \\
\hline
\end{tabular}
\end{table}
\end{center}

\section{Four-quark stability and threshold determination}
\label{thres}

The color degree of freedom makes an important difference between
four-quark systems and ordinary baryons or mesons. For baryons and
mesons it is not possible to construct a color singlet using a
subset of the constituents, thus only $q\bar q$ or $qqq$ states are
proper solutions of the two- or three-quark interacting hamiltonian
and therefore, all solutions correspond to bound states. However,
this is not the case for four-quark systems. The color rearrangement
of Equations~\eref{qq13} and~\eref{qq14} makes that two isolated
mesons are also a solution of the four-quark hamiltonian. In order
to distinguish between four-quark bound states and simple pieces of
the meson-meson continuum, one has to analyze the two-meson states
that constitute the threshold for each set of quantum numbers.

These thresholds must be determined assuming quantum number
conservation within exactly the same model scheme (same parameters
and interactions) used in the four-quark calculation. When dealing
with strongly interacting particles, the two-meson states should
have well defined total angular momentum ($J$) and parity ($P$), the
{\it coupled} scheme. If two identical mesons are considered, the
spin-statistics theorem imposes a properly symmetrized wave
function. Moreover, $C-$parity should be conserved in the final
two-meson state for those four-quark states with well-defined
$C-$parity. If noncentral forces are not considered, orbital angular
momentum ($L$) and total spin ($S$) are also good quantum numbers,
being this the {\it uncoupled} scheme.

An important property of four-quark states containing identical
quarks, like for instance the $QQ\bar n\bar n$ system, that is
crucial for the possible existence of bound states, is that only one
physical threshold is allowed, $(Q \bar n)(Q\bar n)$ for the case of
heavy-light tetraquarks. Consequently, particular modifications of
the four-quark interaction, for instance a strong color-dependent
attraction in the $QQ$ pair, would not be translated into the
asymptotically free two-meson state. As discussed in~\cite{Vij09b},
this is not a general property of four-quark spectroscopy, since the
$Q\bar Q n\bar n$ four-quark state has two allowed physical
thresholds: $(Q\bar Q)(n\bar n)$ and $(Q\bar n)(n\bar Q)$. The
lowest thresholds for $nn\bar Q\bar Q$ states are given
in~\cite{Vij09b}, for $nQ\bar n\bar Q$ states in~\cite{Vij07}, and
those for $QQ\bar Q\bar Q$ in~\cite{Vij06b}. We give in
Table~\ref*{threstab} the lowest threshold for same particular cases
to illustrate their differences. We show both the coupled (CO) and
the uncoupled (UN) schemes together with the final state relative
orbital angular momentum of the decay products. We would like to
emphasize that even when only central forces are considered the
coupled scheme is the relevant one for experimental observations.

The relevant quantity for analyzing the stability of any four-quark
state is $\Delta_E$, the energy difference between the mass of the
four-quark system and that of the lowest two-meson threshold,
\begin{equation}
\label{binding}
\Delta_E=E_{4q}-E(M_1,M_2)\, ,
\end{equation}
where $E_{4q}$ stands for the four-quark energy and $E(M_1,M_2)$ for
the energy of the two-meson threshold. Thus, $\Delta_E<0$ indicates
that all fall-apart decays are forbidden, and therefore one has a
proper bound state. $\Delta_E\ge 0$ will indicate that the
four-quark solution corresponds to an unbound threshold (two free
mesons).

\section{Probabilities in four-quark systems}
\label{prob}

As discussed in the previous sections four-quark systems present a
richer color structure than ordinary baryons or mesons. While the
color wave function for standard mesons and baryons leads to a
single vector, working with four-quark states there are different
vectors driving to a singlet color state out of colorless or colored
quark-antiquark two-body components. Thus, dealing with four-quark
states an important question is whether we are in front of a
colorless meson-meson molecule or a compact state, i.e., a system
with two-body colored components. While the first structure would be
natural in the naive quark model, the second one would open a new
area on the hadron spectroscopy.

To evaluate the probability of physical channels (singlet-singlet
color states) one needs to expand any hidden-color vector of the
four-quark state color basis in terms of singlet-singlet color
vectors. Given a general four-quark state this requires to mix terms
from two different couplings, Equations~\eref{eq1b} and~\eref{eq1c}.
If $(q_1,q_2)$ or $(\bar q_3,\bar q_4)$ are identical
quarks/antiquarks then, a general four-quark wave function can be
expanded in terms of color singlet-singlet nonorthogonal vectors and
therefore the determination of the probability of physical channels
becomes cumbersome.

In ~\cite{Vij09c} the two Hermitian operators that are well-defined
projectors on the two physical singlet-singlet color states were
derived,
\begin{eqnarray}
{\cal P}_{\ka} & =&  \left( P\hat Q + \hat Q P \right) \frac{1}{2(1-|\paap|^2)}
\nonumber \\
{\cal P}_{\kap} & =&  \left( \hat P Q + Q \hat P \right) \frac{1}{2(1-|\paap|^2)}  \, ,
\label{tt}
\end{eqnarray}
where $P$, $Q$, $\hat P$, and $\hat Q$ are the projectors over the
basis vectors (\ref*{eq1b}) and (\ref*{eq1c}),
\begin{eqnarray}
P & = & \ka \ba \nonumber \\
Q & = & \kb\bb  \, ,
\label{Proj1}
\end{eqnarray}
and
\begin{eqnarray}
\hat P & = & \kap\bap \nonumber \\
\hat Q & = & \kbp\bbp  \, .
\label{Proj2}
\end{eqnarray}

Using them and the formalism of ~\cite{Vij09c}, the four-quark nature (unbound, molecular
or compact) can be explored. Such a formalism can be applied to any four-quark
state, however, it becomes much simpler when distinguishable quarks are present. This would be,
for example, the case of the $nQ\bar n\bar Q$ system, where the Pauli principle does not apply.
In this system the bases \eref{eq1b} and \eref{eq1c} are distinguishable due to the flavor part,
they correspond to $[(n\bar c)(c\bar n)]$ and $[(n\bar n)(c\bar c)]$ as indicated in Table~\ref*{tab_clasi},
and therefore they are orthogonal. This makes that the probability of a physical channel
can be evaluated in the usual way for
orthogonal basis~\cite{Vij08}. The non-orthogonal bases formalism is required for those cases
where the Pauli Principle applies either for the quarks or the antiquarks pairs, see Table~\ref*{tab_clasi}.
Relevant expressions can be found in~\cite{Vij09c}.

\begin{table}[h!!]
\begin{center}
\caption{Mass, in MeV, and flavor dominant component of the
light scalar-isoscalar mesons.}
\label{t8}
\begin{tabular}{|cc|cc|}
\hline
State & PDG & Mass & Flavor \\
\hline
$f_0(600)$            &  400$-$1200        &  568  &  $(n\bar n_{1P})$     \\
$f_0(980)$            &  980$\pm$10        &  999  &  $(nn\bar n\bar n)$   \\
$f_0(1200-1600)$      &  1400$\pm$200      & 1299  &  $(s\bar s_{1P})$     \\
$f_0(1370)$           &  1200$-$1500       & 1406  &  $(n \bar n_{2P})$    \\
$f_0(1500)$           &  1507$\pm$5        & 1611  &  $(ns\bar n\bar s)$   \\
$f_0(1710)$           &  1714$\pm$5        & 1704  &  (glueball)           \\
$f_0(1790)$           &  1790$^{+40}_{-30}$& 1782  &  $(n \bar n_{3P})$    \\
$f_0(2020)$           &  1992$\pm$16       & 1902  &  $(ss \bar s\bar s)$  \\
$f_0(2100)$           &  2103$\pm$17       & 1946  &  $(s \bar s_{2P})$    \\
$f_0(2200)$           &  2197$\pm$17       & 2224  &  $(s \bar s_{3P})$    \\
\hline
\end{tabular}
\end{center}
\end{table}

\section{Some selected results}
\label{results}

To illustrate the formalism we have introduced, we discuss some illustrative results. We
make use of a standard quark potential model, the constituent quark cluster (CQC) model.
It was proposed in the early 90's in an attempt to obtain a simultaneous description of the nucleon-nucleon
interaction and the baryon spectra~\cite{Rep05}.
Later on it was generalized to all flavor sectors giving
a reasonable description of the meson \cite{Vij05a} and baryon spectra~\cite{Vij04a,Vij04b,Vij04c}.
Explicit expressions of the interacting potentials and a detailed discussion of the model can be found in
~\cite{Vij05a}.

The performance of the numerical procedure we have presented described can be checked
by comparing with other methods in the literature to understand its capability and
advantages. Ref.~\cite{Vij09b} makes use of a hyperspherical harmonic (HH) expansion
to study heavy-light tetraquarks, obtaining a mass of
3860.7 MeV ($K_{\rm max}=24$) for the $(L,S,I)=(0,1,0)$ $cc \bar n\bar n$ state
using the CQC model.
The variational formalism described here gives a value of
3861.4 MeV (with 6 Gaussians),  in very good agreement.
Concerning the unbound states, belonging
to the two-meson continuum, the variational is able to describe
reasonably their energies and root mean square radii. For the
unbound $(L,S,I)=(0,0,1)$ $cc \bar n\bar n$ state the variational
method gives a value of $\Delta_E=+5$ MeV to be compared with the
value obtained with the HH formalism ($K=28$), $\Delta_E=+33$. This is
due to the flexibility of the expansion in terms of generalized Gaussians
and its ability to mimic the oscillatory behavior of the continuum wave functions,
something that is more difficult using an expansion in terms of
Laguerre functions~\cite{Vij09b}.
\begin{figure}[h!!]
\begin{center}
\caption{Regge trajectories for the scalar-isoscalar mesons. The squares represent
the results of Table~\protect\ref*{t8}. The lower solid line corresponds to
$n \bar n$ systems and the upper line to $s \bar s$ systems.
The dashed lines correspond to the mass
of those states with a large non$-q\bar q$ component.}
\label{figre}
\vspace*{1.0cm}
\epsfig{file=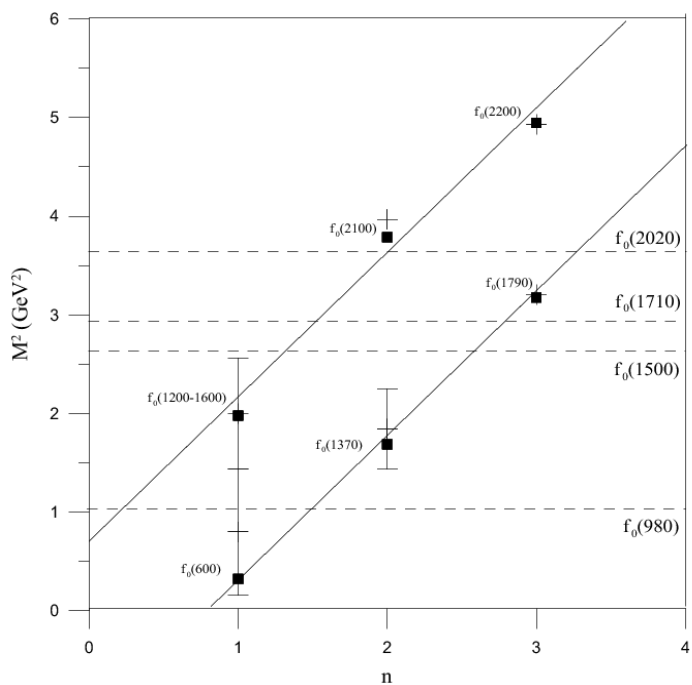,width=3.5in}
\end{center}
\end{figure}

Let us now discussed some particular examples where four-quark
structures could be present. First of all we center
our attention on the light scalar-isoscalar mesons.
In ~\cite{Vij05} scalar mesons below 2 GeV were studied in terms
of the mixing of a chiral nonet of tetraquarks with conventional $q\bar q$ states using the scheme described in
Section~\ref*{mixing}. We show in Table~\ref*{t8} results for the
energies and dominant flavor component of the scalar-isoscalar mesons
when considering also the mixing with a scalar glueball based on intuition from
lattice QCD~\cite{Bal01,Mcn00,Lee00,Ams95}.
The results show a nice correspondence between theoretical
predictions and experiment. This assignment suggests that there are four
isoscalar mesons that are not dominantly
$q \bar q$ states, they are the
$f_0(980)$ (dominantly a $nn \bar n \bar n$  state), the
$f_0(1500)$ (dominantly a $ns \bar n \bar s$  state), the
$f_0(1710)$ (dominantly a glueball) and the
$f_0(2020)$ (dominantly a $ss \bar s \bar s$  state).
This is clearly seen in Figure~\ref*{figre} where we have
constructed the two Regge trajectories associated to
the isoscalar mesons. As it is observed the masses of the
$f_0(600)$, $f_0(1200-1600)$, $f_0(1370)$, $f_0(1790)$, $f_0(2100)$, $f_0(2200)$ fit
nicely in one of the two Regge trajectories, while those
corresponding to the $f_0(980)$, $f_0(1500)$, $f_0(1710)$, $f_0(2020)$ do not
fit for any integer value. The exception would be the $f_0(2020)$
that it is the orthogonal state to the $f_0(2100)$ having
almost 50\% of four-quark component.
The glueball component is shared between the three neighboring states: 20 \% for the $f_0(1370)$,
2 \% for the $f_0(1500)$ and 76 \% for the $f_0(1710)$.
These results assigning the larger glueball component to the $f_0(1710)$ are on
the line with Refs.~\cite{Mcn00,Lee00} and differ from those of
Refs.~\cite{Ams02,Ams96a,Ams96b} concluding that the $f_0(1710)$ is dominantly $s \bar s$
and Ref.~\cite{Ven06} supporting a low-lying glueball camouflaged within the
$f_0(600)$ peak.
\begin{table}[h!!]
\caption{Probabilities (P), in \%, of the wave function components
and masses (QM), in MeV, of the open-charm and open-bottom mesons with $I=0$ (left) and
$I=1/2$ (right) once the mixing between $q\bar q$ and $qq\bar q\bar q$ configurations
is considered. Experimental data (Exp.) are taken from Ref. \cite{PDG08}.}
\label{t3}
\begin{center}
\begin{tabular}{|c|cc||c|cc||c|cc|}
\hline
\multicolumn{6}{|c||}{$I=0$}    & \multicolumn{3}{|c|}{$I=1/2$} \\
\hline
\multicolumn{3}{|c||}{$J^P=0^+$}    & \multicolumn{3}{|c||}{$J^P=1^+$} &
\multicolumn{3}{|c|}{$J^P=0^+$} \\
\hline
QM                      &2339       &2847   &QM         &2421       &2555 &
QM                      &2241               &2713    \\
Exp.                    &2317.8$\pm$0.6 &$-$    &Exp.           &2459.6$\pm$0.6 &$2535.4 \pm 0.6$ &
Exp.            &2352$\pm$50    &$-$\\
\hline
P($cn\bar s\bar n$)     &28         &55 &P($cn\bar s\bar n$)    &25         &$\sim 1$ &
P($cn\bar n\bar n$)     &46                 &49  \\
P($c\bar s_{1^3P}$) &71   &25  &P($c\bar s_{1^1P}$) &74  &$\sim 1$  &
P($c\bar n_{1P}$)       &53                 &46 \\
P($c\bar s_{2^3P}$) &$\sim 1$  &20  &P($c\bar s_{1^3P}$)&$\sim 1$ &98   &
P($c\bar n_{2P}$)       &$\sim 1$           &5 \\
\hline
\hline
QM                  &5679       &6174   &QM         &5713       &5857
&QM                   &5615     &6086 \\
\hline
P($bn\bar s\bar n$) &0.30   &0.51   &P($bn\bar s\bar n$)    &0.24       &$\sim 0.01$
&P($bn\bar n\bar n$)  &0.48       &0.46  \\
P($b\bar s_{1^3P}$) &0.69       &0.26   &P($b\bar s_{1^1P}$)    &0.74       &$\sim 0.01$
&P($b\bar n_{1P}$)    &0.51       &0.47 \\
P($b\bar s_{2^3P}$) &$\sim 0.01$    &0.23   &P($b\bar s_{1^3P}$)    &$\sim 0.01$    &0.99
&P($b\bar n_{2P}$)    &$\sim 0.01$ &0.07 \\
\hline
\end{tabular}
\end{center}
\end{table}

Another interesting scenario where four-quark states may help in the understanding
of the experimental data is the open-charm meson sector~\cite{Vij06,Vij08,Vij09}.
The positive parity open-charm mesons present unexpected properties quite different
from those predicted by quark potential models if a pure $c\bar q$ configuration
is considered. We include in Table~\ref*{t3} some results considering the mixing
between $c\bar q$ configurations and four-quark states.
Let us first analyze the nonstrange sector. The
$^{3}P_{0}$ $c\bar n$ pair and the $cn\bar n\bar n$
have a mass of
2465 MeV and 2505 MeV, respectively. Once the mixing is considered
one obtains a state at 2241 MeV with 46\% of four-quark component
and 53\% of $c\bar n$ pair. The lowest state, representing
the $D^*_0(2308)$, is above the isospin preserving threshold $D\pi$,
being broad as observed experimentally.
The mixed configuration compares much better with
the experimental data than the pure $c\bar n$ state.
The orthogonal state appears higher in energy, at 2713 MeV, with
and important four-quark component.

Concerning the strange sector, the $D_{sJ}^*(2317)$ and the $D_{sJ}(2460)$
are dominantly $c\bar s$ $J=0^+$ and $J=1^+$ states, respectively,
with almost  30\% of four-quark component. Without being dominant,
it is fundamental to shift the mass of the unmixed states to
the experimental values below the $DK$ and $D^*K$ thresholds.
Being both states below their isospin-preserving
two-meson threshold, the only allowed strong decays to
$D_s^* \pi$ would violate isospin and are expected to
have small widths. This width has been estimated assuming either a $q\bar q$ structure
\cite{God03,Bar03}, a four-quark state \cite{Nie05} or vector meson dominance \cite{Col03}
obtaining in all cases a width of the order of 10 keV.
The second isoscalar $J^P=1^+$ state, with an energy of 2555 MeV and
98\% of $c\bar{s}$ component, corresponds to the $D_{s1}(2536)$.
Regarding the $D_{sJ}^*(2317)$, it has been argued that a
possible $DK$ molecule would be preferred with
respect to an $I=0$ $cn\bar s\bar n$ tetraquark,
what would anticipate an $I=1$ $cn\bar s\bar n$ partner
nearby in mass \cite{Barb3}.
The present results support the last argument, namely, the vicinity
of the isoscalar and isovector tetraquarks. However, the coupling
between the four-quark state and the $c\bar s$ system, only allowed
for the $I=0$ four-quark states due to isospin conservation, opens
the possibility of a mixed nature for the $D_{sJ}^*(2317)$, the
remaining $I=1$ pure tetraquark partner appearing much higher in
energy.
The $I=1$ $J=0^{+}$ and $J=1^{+}$ four-quark states appear above
2700 MeV and cannot be shifted to lower energies.
\begin{center}
\begin{table}[h!!]
\begin{center}
\caption{Heavy-light four-quark state properties for selected
quantum numbers. All states have positive parity and total orbital
angular momentum $L=0$. Energies are given in MeV. The notation
$M_1M_2\mid_{\ell}$ stands for mesons $M_1$ and $M_2$ with a
relative orbital angular momentum $\ell$. $P[| \bar 3
3\rangle_c^{12}(| 6\bar 6\rangle_c^{12})]$ stands for the
probability of the $3\bar 3(\bar 6 6)$ components given in
Equation~(\ref*{eq1a}) and $P[\ka(\kb)]$ for the $11(88)$ components
given in Equation~(\ref*{eq1b}). $P_{MM}$, $P_{MM^*}$, and
$P_{M^*M^*}$ have been calculated following the formalism
of~\cite{Vij09c}, and they represent the probability of finding
two-pseudoscalar ($P_{MM}$), a pseudoscalar and a vector
($P_{MM^*}$) or two vector ($P_{M^*M^*}$) mesons}. \label{re1}
\begin{tabular}{|c|ccccc|}
\hline
$(S,I)$             & (0,1)      &  (1,1)      & (1,0)     & (1,0)      & (0,0) \\
Flavor              &$cc\bar n\bar n$&$cc\bar n\bar n$&$cc\bar n\bar n$&$bb\bar n\bar n$&$bb\bar n\bar n$\\
\hline
Energy              & 3877       &  3952       & 3861      & 10395      & 10948 \\
Threshold           & $DD\mid_S$     &  $DD^*\mid_S$   & $DD^*\mid_S$   & $BB^*\mid_S$   &  $B_1B\mid_P$\\
$\Delta_E$              & +5         &  +15        & $-76$     & $-$217     &  $-153$ \\
\hline
$P[| \bar 3 3\rangle_c^{12}]$   & 0.333      &  0.333      & 0.881     & 0.974      &  0.981 \\
$P[| 6 \bar 6\rangle_c^{12}]$   & 0.667      &  0.667      & 0.119     & 0.026      &  0.019 \\
\hline
$P[\ka]$            & 0.556      &  0.556      & 0.374     & 0.342      &  0.340 \\
$P[\kb]$            & 0.444      &  0.444      & 0.626     & 0.658      &  0.660 \\
\hline
$P_{MM}$            & 1.000      &  $-$        & $-$       & $-$        &  0.254 \\
$P_{MM^*}$          & $-$        &  1.000      & 0.505     & 0.531      &  $-$ \\
$P_{M^*M^*}$            & 0.000      &  0.000      & 0.495     & 0.469      &  0.746 \\
\hline
\end{tabular}
\end{center}
\end{table}
\end{center}

We finally tackled an interesting problem in tetraquark
spectroscopy, the molecular or compact nature of four-quark bound
states. This problem requires the determination of probabilities in
non-orthogonal bases mathematically addressed in ~\cite{Vij09c}. We
show in Table~\ref*{re1} some examples of results obtained for
heavy-light tetraquarks. One can see how independently of their
binding energy, all of them present a sizable octet-octet component
when the wave function is expressed in the~\eref{eq1b} coupling. Let
us first of all concentrate on the two unbound states, $\Delta_E >
0$, one with $S=0$ and one with $S=1$, given in Table~\ref*{re1}. The
octet-octet component of basis~\eref{eq1b} can be expanded in terms
of the vectors of basis~\eref{eq1c} as explained in the previous
section. Then, the probabilities are concentrated into a single
physical channel, $MM$ or $MM^*$ [$MM$ stands for two identical
pseudoscalar $D$ ($B$) mesons and $MM^*$ for a pseudoscalar $D$
($B$) meson together with its corresponding vector excitation, $D^*$
($B^*$)]. In other words, the octet-octet component of the
basis~\eref{eq1b} or~\eref{eq1c} is a consequence of having
identical quarks and antiquarks. Thus, four-quark unbound states are
represented by two isolated mesons. This conclusion is strengthened
when studying the root mean square radii, leading to a picture where
the two quarks and the two antiquarks are far away, $\langle
x^2\rangle^{1/2}\gg 1$ fm and $\langle y^2\rangle^{1/2}\gg 1$ fm,
whereas the quark-antiquark pairs are located at a typical distance
for a meson, $\langle z^2\rangle^{1/2}\le 1$ fm. Let us now turn to
the bound states shown in Table~\ref*{re1}, $\Delta_E < 0$, one in
the charm sector and two in the bottom one. In contrast to the
results obtained for unbound states, when the octet-octet component
of basis~\eref{eq1b} is expanded in terms of the vectors of
basis~\eref{eq1c}, one obtains a picture where the probabilities in
all allowed physical channels are relevant. It is clear that the
bound state must be generated by an interaction that it is not
present in the asymptotic channel, sequestering probability from a
single singlet-singlet color vector from the interaction between
color octets. Such systems are clear examples of compact four-quark
states, in other words, they cannot be expressed in terms of a
single physical channel.
\begin{figure}[h!!]
\begin{center}
\caption{\label{f2}$P_{MM}$ as a function of $\Delta_E$.}
\epsfig{file=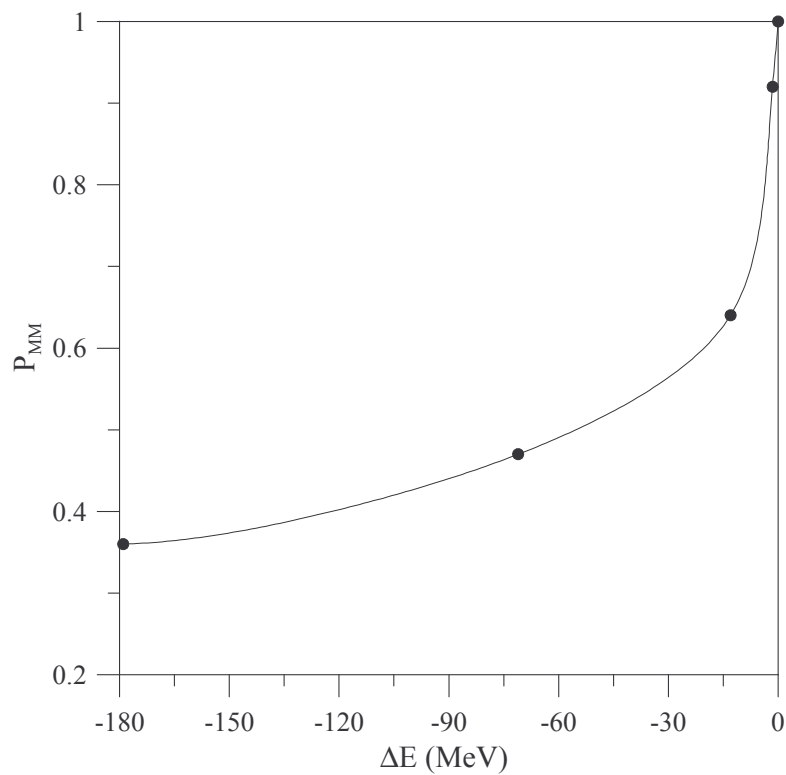,width=4in}
\end{center}
\vspace*{-0.5cm}
\label{fignew}
\end{figure}

We have studied the dependence of the probability of a physical channel
on the binding energy. For this purpose we have considered the simplest
system from the numerical point of view, the
$(S,I)=(0,1)$ $cc\bar n\bar n$ state. Unfortunately, this state
is unbound for any reasonable set of parameters. Therefore, we bind it by multiplying the
interaction between the light quarks by a fudge factor.
Such a modification does not affect the two-meson threshold while
it decreases the mass of the four-quark state. The results are illustrated in
Figure~\ref*{fignew}, showing how in the $\Delta_E\to0$ limit,
the four-quark wave function is almost a pure single physical
channel. Close to this limit one would find what could be defined as
molecular states. When the probability concentrates
into a single physical channel ($P_{M_1M_2}\to 1$) the
system gets larger than two isolated mesons~\cite{Vij09}.
One can identify the subsystems responsible for increasing the size of the four-quark state.
Quark-quark ($\langle x^2\rangle^{1/2}$) and antiquark-antiquark ($\langle y^2\rangle^{1/2}$)
distances grow rapidly while the quark-antiquark  distance ($\langle z^2\rangle^{1/2}$)
remains almost constant. This reinforces our previous result, pointing to the appearance
of two-meson-like structures whenever the binding energy goes to zero.

\section{Summary}
\label{summary}

We have presented a detailed analysis of the symmetry properties of a four-quark wave
function and its solution by means of a variational approach for simple
Hamiltonians. The numerical capability of the method has been analyzed.
We have also emphasized the relevance of a correct analysis of the two-meson
thresholds when dealing with the stability of four-quark systems.
We have discussed the potential importance of four-quark structures
in several different systems: the light scalar-isoscalar mesons and the
open-charm mesons. We have also introduced the necessary ingredients
to study the nature of four-quark bound states, distinguishing
between molecular and compact four-quark states.

Although the present analysis has been performed by means of a particular
quark interacting potential, the CQC model, the conclusions derived are independent
of the quark-quark interaction
used. They mainly rely on using the same hamiltonian to describe tensors of
different order, two and four-quark components in the present case. When dealing with a
complete basis, any four-quark deeply bound state has to be compact. Only
slightly bound systems could be considered as molecular. Unbound states correspond
to a two-meson system. A similar situation would
be found in the two baryon system, the deuteron could be considered as a
molecular-like state with a small percentage of its wave function on the $\Delta \Delta$ channel,
whereas the $H-$dibaryon would be a compact six-quark state.
When working with central forces, the only way of getting a bound system is to have
a strong interaction between the constituents that are far apart in the asymptotic limit
(quarks or antiquarks in the present case). In this case the short-range
interaction will capture part of the probability of a two-meson threshold to form a bound
state. This can be reinterpreted as an infinite sum over physical states.
This is why
the analysis performed here is so important before any conclusion can be made concerning
the existence of compact four-quark states beyond simple molecular structures.

If the prescription of using the same hamiltonian to describe all tensors in the Fock space is relaxed,
new scenarios may appear. Among them, the inclusion of many-body forces is particularly relevant.
In ~\cite{Vij07ba,Vij07bb} the stability of $QQ\bar n\bar n$ and $Q\bar Q n \bar n$ systems
was analyzed in a simple string model considering only a multiquark confining interaction given
by the minimum of a flip-flop or a butterfly potential in an attempt to discern whether
confining interactions not factorizable as two-body potentials would influence the stability
of four-quark states. The ground state of systems made of two quarks and two antiquarks of
equal masses was found to be below the dissociation threshold. While for the cryptoexotic
$Q\bar Q n\bar n$ the binding decreases when increasing the mass ratio $m_Q/m_n$, for the
flavor exotic $QQ\bar n\bar n$ the effect of mass symmetry breaking is opposite. Others scenarios may emerge
if different many-body forces, like many-body color interactions~\cite{Dmi01a,Dmi01b} or 't Hooft
instanton-based three-body interactions~\cite{Hoo76}, are considered.

\section{Acknowledgements}
This work has been partially funded by Ministerio de Ciencia y Tecnolog\'{\i}a
under Contract No. FPA2007-65748 and by EU FEDER, by Junta de Castilla y Le\'{o}n
under Contracts No. SA016A17 and GR12,
by the Spanish Consolider-Ingenio 2010 Program CPAN (CSD2007-00042),
by HadronPhysics2, a FP7-Integrating Activities and Infrastructure
Program of the European Commission, under Grant 227431, and by
Generalitat Valenciana, PROMETEO/2009/129.

\bibliographystyle{mdpi}


\begin{thebibliography}{11}


\bibitem{Ge64} Gell-Mann, M. A schematic model of baryons and mesons.
        {\em Phys. Lett.} {\bf 1964}, {\em 8}, 214--215.

\bibitem{Ja77} Jaffe, R.~L. Multiquark hadrons.I.Phenomenology of $Q^2{\overline Q}^2$ mesons.
        {\em Phys. Rev. D} {\bf 1977}, {\em 15}, 267--280.

\bibitem{Mo96} Moinester, M.~A. How to search for doubly charmed baryons and tetraquarks.
        {\em Z. Phys. A} {\bf 1996}, {\em 355}, 349--362.

\bibitem{ZS86a} Zouzou, S.; Silvestre-Brac, B.; Gignoux, C.; Richard, J.~M. Four quark bound states.
        {\em Z. Phys. C} {\bf 1986}, {\em 30}, 457--468.
\bibitem{ZS86b} Heller, L.;  Tjon, J.~A. On the existence of stable dimesons.
        {\em Phys. Rev. D}, {\bf 1987}, {\em 35}, 969--974.

\bibitem{PDG08} Amsler, C.; \textit{et al.} Review of Particle Physics.
        {\em Phys. Lett. B }, {\bf 2008}, {\em 667}, 1--1340.

\bibitem{Jaf05a} Jaffe, R.~L. Exotica.
        {\em Phys. Rep.} {\bf 2005}, {\em 409}, 1--45.
\bibitem{Jaf05b} Swanson, E.~S. The new heavy mesons: A status report.
        {\em Phys. Rep.} {\bf 2006}, {\em 429}, 243--305.
\bibitem{Jaf05c}  Klempt, E.; Zaitsev, A. Glueballs, hybrids, multiquarks: Experimental facts
        versus QCD inspired concepts.
        {\em Phys. Rep.} {\bf 2007}, {\em 454}, 1--202.

\bibitem{Ade82a} Ader, J.~P.; Richard, J.~M.; Taxil, P. Do narrow heavy multiquark states exist?
                {\em Phys. Rev. D} {\bf 1982}, {\em 25}, 2370--2382.
\bibitem{Ade82b} Ballot, J.~L.; Richard, J.~M. Four quark states in additive potentials.
                {\em Phys. Lett. B} {\bf 1983}, {\em 123}, 449--451.

\bibitem{Vij07ba} Vijande, J.; Valcarce, A.; Richard, J.-M. Stability of multiquarks in a simple string model.
        {\em Phys. Rev. D} {\bf 2007}, {\em 76}, 114013:1--114013:5.
\bibitem{Vij07bb}
        Ay, C.; Richard, J.~ M.; Rubinstein, J.~H. Stability of asymmetric tetraquarks in the
        minimal-path linear potential.
                Phys. Lett. B {\bf 2009}, {\em 674}, 227--231.

\bibitem{Va88} Varshalovich, D.~A.; Moskalev, A.~N.; Khersonsky, V.~K. Quantum theory of angular momentum.
        Eds: World Scientific, {\bf 1988}.

\bibitem{De63a} de Swart, J.~J. The octet model and its Clebsch-Gordan coefficients.
        {\em  Rev. Mod. Phys.} {\bf 1963}, {\em 35}, 916--939.
\bibitem{De63b} Kaeding, T.~A. Tables of SU(3) isoscalar factors.
        {\em nucl-th/9502037} {\bf 1995}.

\bibitem{Vij06} Vijande, J.; Fern\'andez, F.; Valcarce, A. Open-charm meson spectroscopy.
        {\em Phys. Rev. D} {\bf 2006} ,{\em 73}, 034002:1--034002:9.

\bibitem{Vij05} Vijande, J.; Valcarce, A.; Fern\'andez, F.; Silvestre-Brac, B. Nature of the light scalar mesons.
        {\em Phys. Rev. D} {\bf 2005}, {\em 72}, 034025:1--034025:8.

\bibitem{Vij08} Vijande, J.; Valcarce, A.; Fern\'andez, F. B meson spectroscopy.
        {\em Phys. Rev. D} {\bf 2008}, {\em 77}, 017501:1--017501:4.

\bibitem{Vij09} Vijande, J.; Valcarce, A.; Fern\'andez, F. A multiquark description
        of the $D_{sJ}(2860)$ and $D_{sJ}(2700)$.
        {\em Phys. Rev. D} {\bf 2009}, {\em 79}, 037501:1--037501:4.

\bibitem{Vij09b} Vijande, J.; Valcarce, A.; Barnea, N. Exotic meson-meson molecules and compact four--quark states.
        {\em Phys. Rev. D} {\bf 2009}, {\em 79}, 074010:1--074010:16.

\bibitem{Vij07} Vijande, J.; Weissman, E.; Barnea, N.; Valcarce, A. Do $c \bar c n \bar n$ bound states exist?
        {\em Phys. Rev. D} {\bf 2007}, {\em 76}, 094022:1--094022:17).

\bibitem{Vij06b} Barnea, N.; Vijande, J.; Valcarce, A. Four-quark spectroscopy within the
        hyperspherical formalism.
        {\em Phys. Rev. D} {\bf 2006}, {\em 73}, 054004:1-054004:11.

\bibitem{Vij09c} Vijande, J.; Valcarce, A. Probabilities in nonorthogonal bases: Four--quark systems.
        {\em Phys. Rev. C} {\bf 2009}, {\em 80}, 035204:1--035204:10.

\bibitem{Rep05} Valcarce, A.; Garcilazo, H.; Fern\'andez, F.; Gonz\'alez, P.
        Quark--model study of few--baryon systems.
                {\em Rep. Prog. Phys.} {\bf 2005}, {\em 68}, 965--1042.

\bibitem{Vij05a} Vijande, J.; Fern\'andez, F.; Valcarce, A. Constituent quark model study of the meson spectra.
                {\em J. Phys. G} {\bf 2005}, {\em 31}, 481--506.

\bibitem{Vij04a} Valcarce, A.; Garcilazo, H.; Vijande, J. Constituent quark model
        study of light- and strange-baryon spectra.
                {\em Phys. Rev. C} {\bf 2005}, {\em 72}, 025206:1--025206:9.
\bibitem{Vij04b} Garcilazo, H.; Vijande, J.; Valcarce, A. Faddeev study of heavy-baryon spectroscopy.
                {\em J. Phys. G} {\bf 2007}, {\em 34}, 961--976.
\bibitem{Vij04c} Valcarce, A.;Garcilazo, H.; Vijande, J. Towards an understanding of heavy baryon spectroscopy.
        {\em Eur. Phys. J. A} {\bf 2008}, {\em 37}, 217--225.

\bibitem{Bal01} Bali, G.~S. QCD forces and heavy quark bound states.
                {\em Phys. Rep.} {\bf 2001}, {\em 343}, 1--136.

\bibitem{Mcn00} McNeile, C.; Michael, C. Mixing of scalar glueballs and flavor-singlet scalar mesons.
                {\em Phys. Rev. D} {\bf 2001}, {\em 63}, 114503:1--114503:11.

\bibitem{Lee00} Lee, W.; Weingarten, D. Scalar quarkonium masses and mixing with the lightest scalar glueball.
                {\em Phys. Rev. D} {\bf 2000}, {\em 61}, 014015:1--014015:17.

\bibitem{Ams95} Amsler, C.; Close, F.~E. Evidence for a scalar glueball.
                {\em Phys. Lett. B} {\bf 1995}, {\em 353}, 385--390.

\bibitem{Ams02} Amsler, C. Further evidence for a large glue component in the $f_0(1500)$ meson.
                {\em Phys. Lett. B} {\bf 2002}, {\em 541}, 22--28.

\bibitem{Ams96a} Amsler, C.; Close, F.~E. Is $f_0(1500)$ a scalar glueball?
                {\em Phys. Rev. D} {\bf 1996}, {\em 53}, 295--311.
\bibitem{Ams96b} Close, F.~E.; Kirk, A. Scalar glueball$-q\bar q$ mixing above 1 GeV and implications for lattice QCD.
                {\em Eur. Phys. J. C} {\bf 2001}, {\em 21}, 531--543.

\bibitem{Ven06} Vento, V. Scalar glueball spectrum.
                {\em Phys. Rev. D} {\bf 2006}, {\em 73}, 054006:1--054006:7.

\bibitem{God03} Godfrey, S. Testing the nature of the $D_{sJ}^*(2317)^+$ and $D_{sJ}(2463)^+$
        states using radiative transitions.
                {\em Phys. Lett. B} {\bf 2003}, {\em 568}, 254--260.

\bibitem{Bar03} Bardeen, W.~A.; Eichten, E.~J.; Hill, C.~T. Chiral multiplets of heavy-light mesons.
                {\em Phys. Rev. D} {\bf 2003}, {\em 68}, 054024:1--054024:11.

\bibitem{Nie05} Nielsen, M. $D^+_{sJ}(2317) \to D^+_s \pi^0$ decay width.
                {\em Phys. Lett. B} {\bf 2006}, {\em 634}, 35--38.

\bibitem{Col03} Colangelo, P.; de Fazio, F. Understanding $D_{sJ}(2317)$.
                {\em Phys. Lett. B} {\bf 2003}, {\em 570}, 180--184.

\bibitem{Barb3} Barnes, T.; Close, F.~E.; Lipkin, H.~J. Implications of a DK molecule at 2.32 GeV.
                {\em Phys. Rev. D} {\bf 2003}, {\em 68}, 054006 (1--5).

\bibitem{Dmi01a} Dmitra\v{s}inovi\'c, V. Cubic Casimir operator of $SU_C(3)$ and confinement
        in the nonrelativistic quark model.
                {\em Phys. Lett. B} {\bf 2001}, {\em 499}, 135--140.
\bibitem{Dmi01b} Dmitra\v{s}inovi\'c, V. Color SU(3) symmetry, confinement, stability, and
        clustering in the $q^2q^2$ system.
                {\em Phys. Rev. D} {\bf 2003}, {\em 67}, 114007:1--114007:12.

\bibitem{Hoo76} 't Hooft, G. Computation of the quantum effects due to a four-dimensional pseudoparticle.
                {\em Phys. Rev. D} {\bf 1976}, {\em 14}, 3432--3450. (Erratum
                {\em Phys. Rev. D} {\bf 1978}, {\em 18}, 2199--2200.)

\end{thebibliography}

\end{document}